\documentclass[10pt,fleqn,a4paper,twoside]{article}

\usepackage[margin=1in]{geometry}
\usepackage[justification=centering]{caption}
\usepackage{amsmath}
\usepackage{authblk}
\usepackage[figuresright]{rotating}
\usepackage[english]{babel}

\begin{document}

\title{STABILITY AND ORDER OF ACCURACY ANALYSIS OF HIGH-ORDER SCHEMES FORMULATED USING THE FLUX RECONSTRUCTION APPROACH}

\author[1]{Frederico Bolsoni Oliveira \thanks{Corresponding author: fredericobolsoni@gmail.com}}
\affil[1]{\small{Instituto Tecnológico de Aeronáutica -- DCTA\slash ITA, 12228-900 -- São José dos Campos, SP, Brazil}}

\author[2]{João Luiz F. Azevedo \thanks{joaoluiz.azevedo@gmail.com}}
\affil[2]{\small{Instituto de Aeronáutica e Espaço -- DCTA\slash IAE\slash ACE-L, 12228-904 -- São José dos Campos, SP, Brazil}}

\date{}

\maketitle

\begin{abstract}
    A stability analysis is performed on high-order schemes formulated using the Flux Reconstruction (FR) approach. The one-dimensional advection model equation is used for the assessment of the stability region of these schemes when coupled with Runge-Kutta-type time-march procedures. Schemes are created using different numbers of internal points for each cell of the discrete domain, in such a way that a broad spectrum of spatial discretization orders can be analyzed. Multiple correction functions are employed so that well-known schemes, for instance the Discontinuous Galerkin (DG) and the Staggered-Grid schemes, may also be included in the study. The stability analysis is performed by identifying the behavior, in the complex plane, of the eigenvalues associated with each one of the considered cases. It is observed that, as the order of the scheme increases, a significant decrease in its domain of stability occurs, followed by a significant reduction in the amount of artificial dissipation that is introduced to the solution.
    
    \textbf{Keywords:} \textit{Computational Fluid Dynamics}, \textit{High-Order Schemes}, \textit{Flux Reconstruction}, \textit{Stability Analysis}, \textit{Advection Equation}
\end{abstract}

\section{INTRODUCTION}\label{Sec:Introduction}

In the field of Computational Fluid Dynamics, an important step that has to be taken in order to discretely solve a system of partial differential equations that model the dynamics of fluid flow, at any level of approximation, is to choose an appropriate numerical scheme. Among the properties of the scheme that must be considered when making this decision is the order of the spatial discretization error. Second-order schemes are usually the working-force of most practical problems in the field of aerospace engineering. Well-known, second-order schemes are capable of providing good results for a wide range of problems, as long as a refined enough mesh is used, without compromising the overall robustness of the method.

There are, however, specific cases that are particularly sensible to the presence of artificial dissipation. In such instances, the usage of second-order schemes might be improper. Depending on the computational resources available, the degree of mesh refinement required to keep the artificial dissipation below a certain threshold can become either impractical or completely prohibitive. Cases such as the analysis of low-amplitude acoustic waves and the study of high-lift wing configurations are examples that are heavily affected by this phenomenon \cite{breviglieri2016}. One possible way to either solve or mitigate the introduction of excessive amounts of artificial dissipation in the solution is to use schemes that are of third-order and above. Methods of this class are known as ``high-order schemes'' \cite{wang2012} and have been the subject of intensive research for several decades, gaining traction with the increase in the available computational power of current machines. Besides being capable of achieving lower levels of artificial dissipation when compared to second-order schemes, there are arguments in the literature in favor of using high-order schemes with the intention of achieving similar results to their lower-order counterparts \cite{wang2007}. This is accomplished by using high-order schemes coupled with coarse meshes, which are capable of achieving error levels that are comparable to those observed when second-order schemes are used with fine meshes, but with the advantage of usually being computationally cheaper to execute \cite{wang2007}.

In the context of high-order schemes, mathematical frameworks have been developed in order to construct schemes of arbitrary spatial discretization order \cite{ekaterinaris2005, huynh2014}. Among the well-documented ones are the Discontinuous Galerkin \cite{cockburn1991, bassi1997}, the Spectral Differences \cite{liu2006} and the Spectral Volumes \cite{sun2006} class of schemes. More recently, another approach for constructing high-order schemes has been introduced in the literature: the Flux Reconstruction \cite{huynh2007}, also referred to as FR. This framework is capable of creating compact high-order schemes for solving partial differential equations. Special focus was given to the treatment of advection terms \cite{huynh2007} and diffusion terms \cite{huynh2009}, which are integral components of high-fidelity mathematical models used in fluid dynamics, such as the Navier-Stokes equations.

In the Flux Reconstruction approach, the solution is known at multiple discrete points, also known as nodes, within a discrete domain element: a cell. A continuous solution is reconstructed within the cell by performing an interpolation using a basis of Lagrange polynomials. In general, the overall resulting solution will be continuous only within a cell, usually displaying discontinuities across the interfaces between two adjacent cells. Therefore, if nothing else is done, there is no interaction between nearby cells. The main idea behind the FR framework is to introduce means for information to propagate through the domain. This is done by defining common values for the property fluxes (and its derivatives, if needed) across a cell interface. For the advection terms, the interface fluxes are usually reconstructed in an upwind manner, with the Roe flux \cite{roe1986} being a popular procedure to be used. For the diffusion terms, the common fluxes are taken to be, in its most general case, a weighted average between the fluxes of the immediate adjacent cells. A corrected, continuous, flux function can, then, be reconstructed within a cell, in such a way that the previously defined common interface values are respected. Finally, the time-derivatives of the solution properties can be evaluated by using the corrected fluxes and, then, integrated by using an appropriate time-march scheme selected by the user.

An interesting characteristic of the FR framework is that the continuous reconstruction of the flux terms is performed by using a special set of functions, called ``correction functions'', and, depending on the function(s) used, different high-order schemes are achieved. For instance, schemes such as the nodal Discontinuous Galerkin \cite{cockburn1991} and the Staggered Grid \cite{kopriva1996} can be recovered by using the Flux Reconstruction in combination with an appropriate correction function. One important step that must be made towards the comprehension of the properties of a scheme is the assessment of its stability bounds and effective order of accuracy. In the case of the FR framework, part of this process has already been done by Huynh \cite{huynh2007} and documented for a limited number of scheme orders. The present paper aims to expand this analysis by using the FR framework with 5 different correction functions and using cells with up to 10 internal nodes. The resulting schemes are, then, used to solve the 1-D advection model equation. Observations are done regarding the effects of the artificial dissipation over the transient solution, obtained by using a forth-order Runge-Kutta time-marching procedure (RK4). Further insights regarding the stability and accuracy of the resulting schemes are obtained by performing a Fourier analysis on each one of them.

\section{NUMERICAL FORMULATION}

In this section, a brief explanation is given regarding the Flux Reconstruction approach for the construction of high-order schemes. The intention is to give the reader sufficient background knowledge regarding the inner workings of the FR approach before performing the Fourier stability analysis in the next section.

\begin{figure}[h!]
\centering
\includegraphics[angle=0, width=\textwidth]{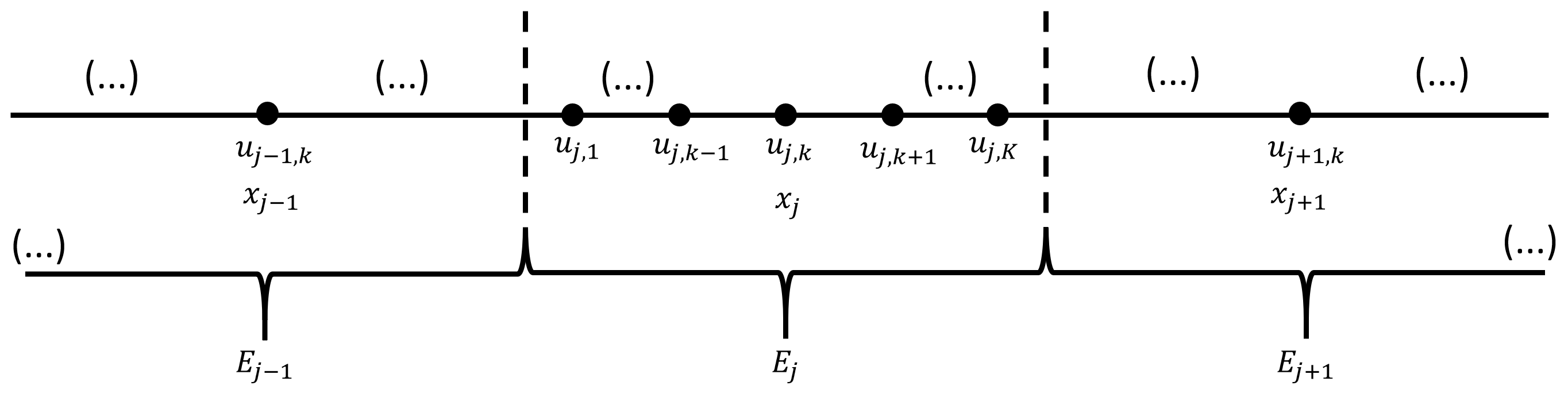}
\caption{Schematic diagram of the discrete domain.}
\label{Fig:DiscreteDomainSchematic}
\end{figure}

The equation of interest is the initial value problem (IVP) given by the 1-D advection model equation:

\begin{equation}  \label{Eq:AdvectionModel}
\begin{cases}
\frac{\partial u}{\partial t} + \frac{\partial f}{\partial x} = 0, \; \; \; \text{ with } \; \; \; f = au \\
u(x,t)|_{t = 0} = u_0(x)
\end{cases}
\end{equation}

\noindent
where $t$ is the time coordinate, $x$ is the space coordinate, $u(x,t)$ is the property being transported by a wave, whose dynamics is given by Eq.~(\ref{Eq:AdvectionModel}), with constant speed $a$. Finally, $f$ is the flux term and the main subject of interest to the FR approach. The discrete domain can be constructed by dividing its continuous counterpart into multiple cells $E$, where the $j$-th cell is denoted by $E_j$ and has a length $h_j$. Each cell is composed of $K$ nodes, which is where the discrete solution, $u(x,t)$, is evaluated at. The centroid of cell $E_j$ is located at $x_j$, and the coordinate of the $k$-th node of the $j$-th cell is denoted by $x_{j,k}$. Following the same naming convention, the solution evaluated at $x_{j,k}$ is defined as $u_{j,k}$, which is a function of time only. The coordinate of the interface located between elements $E_j$ and $E_{j+1}$ is $x_{j+\frac{1}{2}}$. Likewise, $x_{j-\frac{1}{2}}$ is the coordinate of the interface located between $E_{j-1}$ and $E_j$. Figure~\ref{Fig:DiscreteDomainSchematic} shows a schematic of this domain discretization.

Since each cell can have an arbitrary length, $h_j$, and the node disposition within each one of them can also be arbitrary, it is of interest to perform a domain transformation to a local system of coordinates, $\xi(x)$. This is done in such a way that the transformed cell has a constant length of $2$ units, and the left and right boundaries are located at $\xi=-1$ and $\xi=1$, respectively. Here, this transformation is accomplished by using a linear function mapping

\begin{equation} \label{Eq:DomainTransformationMappingXi}
\xi(x) = \frac{2}{h_j} \left( x - x_j \right),
\end{equation}

\begin{equation} \label{Eq:DomainTransformationMappingX}
x(\xi) = \frac{h_j}{2}\xi + x_j.
\end{equation}

Supposing that the discrete solution $u_{j,k}$ is known for a given time $t$, a continuous solution $u_j(\xi)$ can be reconstructed within cell $E_j$ by using the Lagrange polynomials, $\phi_{j,k}(x)$, as a basis for performing a data interpolation

\begin{equation}  \label{Eq:SolutionInterpolation}
u_j(\xi) = \sum_{k=1}^{K} \left[ u_{j,k} \phi_{j,k} (\xi) \right],
\end{equation}

\noindent
where

\begin{equation}  \label{Eq:LagrangePolynomial}
\phi_{j,k}(\xi) = \prod_{l=1, l \neq k}^K \left( \frac{\xi - \xi_{j,l}}{\xi_{j,k} - \xi_{j,l}} \right).
\end{equation}

\noindent
Notice that, from the way that the polynomial basis is constructed, $\phi_{j,k}$ is equal to $1$ at the $k$-th node of $E_j$, and equal to zero at every other node. In a similar manner, a continuous flux function, $f_j(\xi)$, can also be reconstructed from the discrete scalar flux values, $f_{j,k}$, as shown in Eq.~(\ref{Eq:FluxInterpolation}). It must be made clear that the continuous flux function is obtained by interpolating the discrete flux values, as written in Eq.~(\ref{Eq:FluxInterpolation}), and not by evaluating $f$ at the reconstructed solution function $u_j(\xi)$. In other words, $f_j(\xi) \neq f(u_j({\xi}))$ in general.

\begin{equation} \label{Eq:FluxInterpolation}
f_j(\xi) = \sum_{k=1}^{K} \left[ f_{j,k} \phi_{j,k} (\xi) \right].
\end{equation}

\noindent
The derivative of $f_j(\xi)$ can, then, be easily computed by simply deriving the Lagrange polynomials. Hence, the derivatives of the fluxes evaluated at each cell node are simply

\begin{equation} \label{Eq:FluxDerivative}
\frac{d \left( f_j \right)}{d \xi} \biggm| _{\xi_k} \equiv \left( f_\xi \right)_{j,k} =  \sum_{l=1}^{K} \left[ f_{j,k} \frac{d \phi_{j,l}}{d \xi} \biggm|_{\xi_k} \right],
\end{equation}

\begin{equation}
\left( f_x \right)_{j,k} = \frac{2}{h_j} \left( f_\xi \right)_{j,k}.
\end{equation}

\noindent
In tensor notation, Eq.~(\ref{Eq:FluxDerivative}) can be written simply as

\begin{equation} \label{Eq:TensorFluxDerivative}
\overrightarrow{\left( f_\xi \right)_j} = \left[ D \right]_j \overrightarrow{f_j},
\end{equation}

\noindent
where

\begin{equation}  \label{Eq:DiscreteFluxVecNotation}
\overrightarrow{f_j} \equiv
\begin{Bmatrix}
f_{j,1} & f_{j,2} & ... & f_{j,K-1} & f_{j,K}
\end{Bmatrix}^T,
\end{equation}

\begin{equation}
\left[ D \right]_j \equiv
\begin{Bmatrix}
\frac{d \phi_{j,1}}{d \xi} \biggm|_{\xi_1} & \frac{d \phi_{j,2}}{d \xi} \biggm|_{\xi_1} & ... & \frac{d \phi_{j,K-1}}{d \xi} \biggm|_{\xi_1} & \frac{d \phi_{j,K}}{d \xi} \biggm|_{\xi_1} \\
\frac{d \phi_{j,1}}{d \xi} \biggm|_{\xi_2} & \frac{d \phi_{j,2}}{d \xi} \biggm|_{\xi_2} & ... & \frac{d \phi_{j,K-1}}{d \xi} \biggm|_{\xi_2} & \frac{d \phi_{j,K}}{d \xi} \biggm|_{\xi_2} \\
\vdots & \vdots & \ddots & \vdots & \vdots \\
\frac{d \phi_{j,1}}{d \xi} \biggm|_{\xi_K} & \frac{d \phi_{j,2}}{d \xi} \biggm|_{\xi_K} & ... & \frac{d \phi_{j,K-1}}{d \xi} \biggm|_{\xi_K} & \frac{d \phi_{j,K}}{d \xi} \biggm|_{\xi_K} \\
\end{Bmatrix}.
\end{equation}

\noindent
The vector of flux derivatives, $\overrightarrow{\left( f_\xi \right)_j}$, follows an equivalent definition as the one shown in Eq.~(\ref{Eq:DiscreteFluxVecNotation}). Also, it is important to notice that, if the node distribution in local coordinates is the same for all cells, then $\left[ D \right]_j$ is also the same for all cells.

As discussed in the previous section, the internal interpolation of the flux terms does not take into account the interaction that exists between adjacent cells. Therefore, if Eq.~(\ref{Eq:FluxInterpolation}) were inserted directly into Eq.~(\ref{Eq:AdvectionModel}) to compute $\partial u / \partial t$ and perform the time march, the resulting scheme would be unstable, since the characteristic variables related to the solution would be unable to be transported through the domain. Hence, a different definition for the fluxes is required, such that communication between adjacent cells is allowed. This is accomplished by defining a corrected flux function, $F$. The first step in computing it is to set a common value for the flux at each cell interface. A common approach for doing it is to use Roe's upwind flux definition \cite{roe1986}. This upwind flux is constructed based on the mean value theorem, and a simple form of writing it for the right interface, $j+1/2$, in the context of the advection equation is

\begin{equation}  \label{Eq.:RoeFlux}
f_{j+\frac{1}{2}, upwind} = \frac{1}{2} \left[ f \left( u_L \right) + f \left( u_R \right) \right] - \frac{1}{2} \left| a \right| \left( u_R - u_L \right),
\end{equation}

\noindent
in which $u_L$ and $u_R$ are the values of $u$ at the interface, as determined by the cells located directly to the left and to the right of it. That is, $u_L \equiv u_j(1)$ and $u_R \equiv u_{j+1}(-1)$. Equivalent definition can be easily developed for $f_{j-\frac{1}{2}, upwind}$. Notice that, for this problem, $f_{j+\frac{1}{2}, upwind} = a \, u_j(1)$ if $a \geq 0$ and $f_{j+\frac{1}{2}, upwind} = a \, u_{j+1}(-1)$ if $a<0$.

With a common flux value given by Eq.~(\ref{Eq.:RoeFlux}), $F$ can be defined by using two special correction functions, $g_{LB}(\xi)$ and $g_{RB}(\xi)$. These correction functions are constructed in such a way that $g_{LB}(\xi)$ is equal to $1$ at the left interface of the cell, and zero at the right interface. The function $g_{RB}(\xi)$ follows a similar behavior, but related to the right boundary instead. In most cases, including the ones considered here, $g_{RB}(\xi) = g_{LB}(-\xi)$. Hence, only one of them must be defined in order to know both of them. The corrected flux is computed using these correction functions as

\begin{equation}
F_j(\xi) = f_j(\xi) + \left[ f_{j-\frac{1}{2}, upwind} - f_j (-1) \right] g_{LB}(\xi) + \left[ f_{j+\frac{1}{2}, upwind} - f_j (1) \right] g_{RB}(\xi).
\end{equation}

\noindent
The derivative of $F$ can, then, be computed as

\begin{equation}  \label{Eq:CorrectedFluxDerivative}
\frac{d F}{d \xi} \biggm|_{\xi_j}  = \frac{d f}{d \xi} \biggm|_{\xi_j} + \left[ f_{j-\frac{1}{2}, upwind} - f_j (-1) \right] \frac{d \left( g_{LB} \right)}{d \xi} \biggm|_{\xi_j} + \left[ f_{j+\frac{1}{2}, upwind} - f_j(1) \right] \frac{d \left( g_{RB} \right)}{d \xi} \biggm|_{\xi_j},
\end{equation}

\noindent
and the time derivative of the solution is computed from Eq.~(\ref{Eq:AdvectionModel}), as follows

\begin{equation} \label{Eq:DiscreteTimeDerivative}
\left( \frac{\partial u}{\partial t} \right)_{j,k} = - \frac{2}{h_j} \left(\frac{\partial F}{\partial \xi} \right)_{j,k}.
\end{equation}

\noindent
Equation (\ref{Eq:DiscreteTimeDerivative}) is used to perform a time-march by applying an appropriate time integrator. Here, a forth-order, four-stage, Runge-Kutta scheme (RK4) is used \cite{lomax2001}. Since the interpolation of $u$ uses $K$ points, the resulting polynomial is of order $K-1$. In order for the resulting scheme to be concise with this discretization order, the derivative of the corrected flux must also result in a polynomial whose order is of at least $K-1$. Theoretically, the simplest function in which that would be true is a polynomial of order $K$. Hence, the correction functions are taken to be polynomials of $K$-th order.

There are two properties that are still open to be defined. The first one of them is the node distribution within a cell. It is known that, at least for a 1-D linear problem, the internal node disposition does not change the resulting scheme \cite{romero2016}. Thus, for simplicity sake, the internal node distribution is defined to be the Gauss points in the transformed domain \cite{huynh2007}. The second one is the correction function itself. This is exactly what enables the FR framework to construct a wide range of different schemes. In this paper, five different correction functions are considered. The first one of them is $g_1$ (or $g_{DG}$), which uses an adapted form of the Radau polynomial in its definition. The usage of $g_{DG}$ recovers the nodal Discontinuous Galerkin scheme. It is defined as

\begin{equation}
g_{DG_{LB}} = R_{R_K} = \frac{(-1)^K}{2} \left( P_K - P_{K-1} \right),
\end{equation}

\noindent
in which $P_K$ is the $K$-th order Legendre polynomial and $R_{R_K}$ is the $K$-th order modified right Radau polynomial. The zeros of the Legendre polynomials are exactly the Gauss points. In particular, the zeros of the $K$-order Legendre polynomial are the $K$ Gauss points being used here as local node coordinates. Hence, these node coordinates are not zeros of $g_{DG}$.

The second correction function is $g_2$, also known as the Lumped Lobatto correction function, $g_{Lump, Lo}$. It can be expressed as a weighted average of the Radau polynomials

\begin{equation}
g_{Lump, Lo_{LB}} = \frac{K-1}{2K-1} R_{R_K} + \frac{K}{2K-1} R_{R_{K-1}}.
\end{equation}

\noindent
An interesting property of this scheme, not explored here, is the fact that if the internal node coordinates are chosen to be the $K$ Legendre-Lobatto points, then the derivative of this correction function becomes zero at $K-1$ of them, with the exception being the left boundary node, in the case of $g_{LB}$.

The third correction function is $g_{Ga}$, whose $K$ zeros are the right boundary (in the case of $g_{LB}$) and the $K-1$ Gauss points. It can be expressed in a form similar to the previous function

\begin{equation}
g_{Ga_{LB}} = \frac{K}{2K-1} R_{R_K} + \frac{K-1}{2K-1} R_{R_{K-1}}.
\end{equation}

The forth correction function is $g_{Lo}$. It is constructed such that it becomes zero at $K$ of the $K+1$ Legendre-Lobatto points, which includes one of the boundaries (the right one, in the case of $g_{LB}$). If $\xi_{Lo_m}$ are the local coordinates of the $K+1$ Legendre-Lobatto points, then

\begin{equation}
g_{Lo_{LB}} = - \prod_{m=2}^{K+1} \frac{\left( \xi - \xi_{Lo_m} \right)}{1 + \xi_{Lo_m}}.
\end{equation}

Finally, the fifth correction function is $g_{SG}$, which recovers the Staggered-Grid scheme. It uses $K$ of the $K+1$ Chebychev-Lobatto points, which also includes one of the boundaries. If $\xi_{SG_m} = cos((m-1) \pi / (K-1))$ are the local coordinates of the $K+1$ Chebychev-Lobatto points, then

\begin{equation}
g_{SG_{LB}} = - \prod_{m=2}^{K+1} \frac{\left( \xi - \xi_{SG_m} \right)}{1 + \xi_{SG_m}}.
\end{equation}

Therefore, all of the intended schemes have been properly constructed by using the Flux Reconstruction framework.

\section{STABILITY ANALYSIS}

In this section, a general expression for performing the Fourier, or von Neumann \cite{hirschV1}, stability analysis is derived in order to enable the study of the properties of the numerical schemes in the next section.

In a linear Fourier stability analysis, the boundary condition of the partial differential equation at hand is taken to be periodic. Furthermore, the problem is taken to be linear, which is actually not a simplification for the advection model equation. Hence, if the solution is taken to be represented by a complex Fourier series, then each term of the series can be analyzed individually. If the initial conditions of the problem are taken to be periodic in the spatial coordinate, then a single term of its Fourier decomposition can be expressed as $u(x,0)=e^{i \omega x}$, where $\omega$ is the wave-number and $i$ is the imaginary number $\sqrt{-1}$. The analytical solution of this IVP is

\begin{equation}
u_{exact}(x,t) = e^{i \omega (x - at)}.
\end{equation}

If the above expression is used to evaluate the time derivative of the solution, then

\begin{equation}
\frac{\partial u}{\partial t} = -\frac{\partial (au)}{\partial x} \; \; \; \Rightarrow \; \; \; \frac{\partial u_{exact}}{\partial t} = -i\omega \, \left( a u_{exact}(x,t) \right).
\end{equation}

\noindent
After discretizing the domain and the equations, the decoupled system of ordinary differential equations (ODE) tries to mimic the equation shown above. Hence, one of the $K$ eigenvalues, the principal eigenvalue $\lambda_1$, of the discrete matrix operator that acts upon $\overrightarrow{u_j}$ must approximate $-i \omega$, while the other ones are spurious. In order to facilitate the mathematical notation in this section, the following vector variables are defined

\begin{equation}
\vec{\xi} \equiv
\begin{Bmatrix}
\xi_1 & \xi_2 & ... & \xi_K
\end{Bmatrix}^T,
\end{equation}

\begin{equation}
\overrightarrow{g_{\xi}} \equiv
\begin{Bmatrix}
g_{\xi}(\xi_1) & g_{\xi}(\xi_2) & ... & g_{\xi}(\xi_K)
\end{Bmatrix}^T,
\end{equation}

\begin{equation}
\overrightarrow{u_j} \equiv
\begin{Bmatrix}
u_{j,1} & u_{j,2} & ... & u_{j,K}
\end{Bmatrix}^T.
\end{equation}

Supposing that the mesh is homogeneous, that is, $h$ and $\overrightarrow{\xi_j}$ are constants, from Eqs.~(\ref{Eq:DiscreteTimeDerivative}) and (\ref{Eq:CorrectedFluxDerivative}), the following system of ODE's is achieved within a cell

\begin{equation} \label{Eq:IntermediateEq1}
\frac{\partial \overrightarrow{u_j}}{\partial t} = -\frac{2a}{h} \left[ \overrightarrow{u_{\xi_j}} + \left( u_{j-\frac{1}{2}} - u_j(-1) \right) \frac{d \left( g_{LB} \right)}{d \xi} + \left( u_{j+\frac{1}{2}} - u_j(1) \right) \frac{d \left(g_{RB} \right)}{d \xi} \right].
\end{equation}

\noindent
Considering that $a \geq 0$, then Eq.~(\ref{Eq:IntermediateEq1}) reduces to

\begin{equation} \label{Eq:IntermediateEq2}
\frac{\partial \overrightarrow{u_j}}{\partial t} = -\frac{2a}{h} \left[ \overrightarrow{u_{\xi_j}} + \left( u_{j-1}(1) - u_j(-1) \right) \frac{d \left( g_{LB} \right)}{d \xi} \right].
\end{equation}

\noindent
Both $u_{j-1}(1)$ and $u_j(-1)$ can be computed from Eq.~(\ref{Eq:SolutionInterpolation}) by making

\begin{equation} \label{Eq:LeftAndRightVectorOperators}
u_{j}(-1) = \vec{l}^T \overrightarrow{u_j} \; \; \; \text{and} \; \; \; u_{j-1}(1) = \vec{r}^T \overrightarrow{u_j},
\end{equation}

\noindent
with

\begin{equation}
\vec{l} \equiv
\begin{Bmatrix}
\phi_1(-1) & \phi_2(-1) & ... & \phi_K(-1)
\end{Bmatrix}^T,
\end{equation}

\begin{equation}
\vec{r} \equiv
\begin{Bmatrix}
\phi_1(1) & \phi_2(1) & ... & \phi_K(1)
\end{Bmatrix}^T.
\end{equation}

From the considerations made in the Fourier analysis, it is well-known \cite{lomax2001} that $u_{j-1} = e^{-i \omega h} u_j$. Therefore, by using Eqs.~(\ref{Eq:TensorFluxDerivative}) and (\ref{Eq:LeftAndRightVectorOperators}) in Eq.~(\ref{Eq:IntermediateEq2}), followed by a reorganization of the terms, the desired matrix operator form is achieved

\begin{equation} \label{Eq:FullOperatorForm}
\frac{\partial \overrightarrow{u_j}}{\partial t} = \left\{ -\frac{2a}{h} \left[ [D] - \overrightarrow{g_{\xi_{LB}}} \, \vec{l}^T + \left( \overrightarrow{g_{\xi_{RB}}} \, \vec{r}^T \right) e^{-i \omega h} \right] \right\} \overrightarrow{u_j},
\end{equation}

\noindent
or simply

\begin{equation}  \label{Eq:MatrixOperatorForm}
\frac{\partial \overrightarrow{u_j}}{\partial t} = [S] \overrightarrow{u_j},
\end{equation}

\noindent
where the general form of an element of $[S]$ is

\begin{equation}  \label{Eq:SingleElementOperatorForm}
S_{p,q} = -\frac{2a}{h} \left[ \frac{d \left( \phi_q \right)}{d \xi} \biggm|_{\xi_p} + \frac{d \left( g_{LB} \right)}{d \xi} \biggm|_{\xi_p} \left( \phi_q(1) \, e^{-i \omega h} - \phi_q(-1) \right) \right].
\end{equation}

The stability analysis is performed by observing the behavior of the principal eigenvalue, $\lambda_1$, of the matrix operator, $[S]$, with $\omega$ varying from $0$ to $2\pi$. Here, $h$ and $a$ are taken to be equal to $1$. The presence of a positive, real component of $\lambda_1$ for any value of $\omega$ points towards the existence of unstable modes in the scheme, which could lead the solution process to diverge. If these values are positive, but small, the problem only manifests as a mild instability, most likely not being capable of affecting the discrete transient solution if the integration time is not long enough for the unstable modes to develop. On the other hand, the minimum real eigenvalue can be used to compute the CFL (Courant-Friedrichs-Lewy number) limit, associated with a given scheme, when coupled with a known time-march procedure. The overall process of computing this value makes use of the $\sigma$-$\lambda$ relation that exists for each time-marching scheme, as described in Ref.~\cite{lomax2001}, and assumes a limiting value when the real component of $\lambda_1$ is at its minimum.

Finally, the behavior of the principal eigenvalue, for small values of $\omega$, can be used to estimate the effective order of each scheme. The process is described in Ref.~\cite{huynh2007}, and involves measuring how the error that exists between $\lambda_1$ and the analytical solution, $-i\omega$, behaves for small values of $\omega$, which is theoretically the state where the error is still small. This approximation assumes the form

\begin{equation} \label{Eq:ApproxAccuracy}
m \approx \left[ \frac{log_{10} \left( \frac{E(\omega)}{E(\omega / 2)} \right)}{log_{10}(2)} \right] - 1,
\end{equation}

\noindent
where $m$ is the order of the scheme and $E(\omega)$ is the error associated with $\lambda_1$ when the wave-number, $\omega$, is a small arbitrary value. The error term can be computed as

\begin{equation}
E(\omega) = \lambda_1 \biggm|_{\omega} - (- i\omega).
\end{equation}

\section{RESULTS AND DISCUSSION}

\begin{figure}[h!]
    \centering
    \begin{minipage}{0.5\textwidth}
        \centering
        \includegraphics[width=\textwidth]{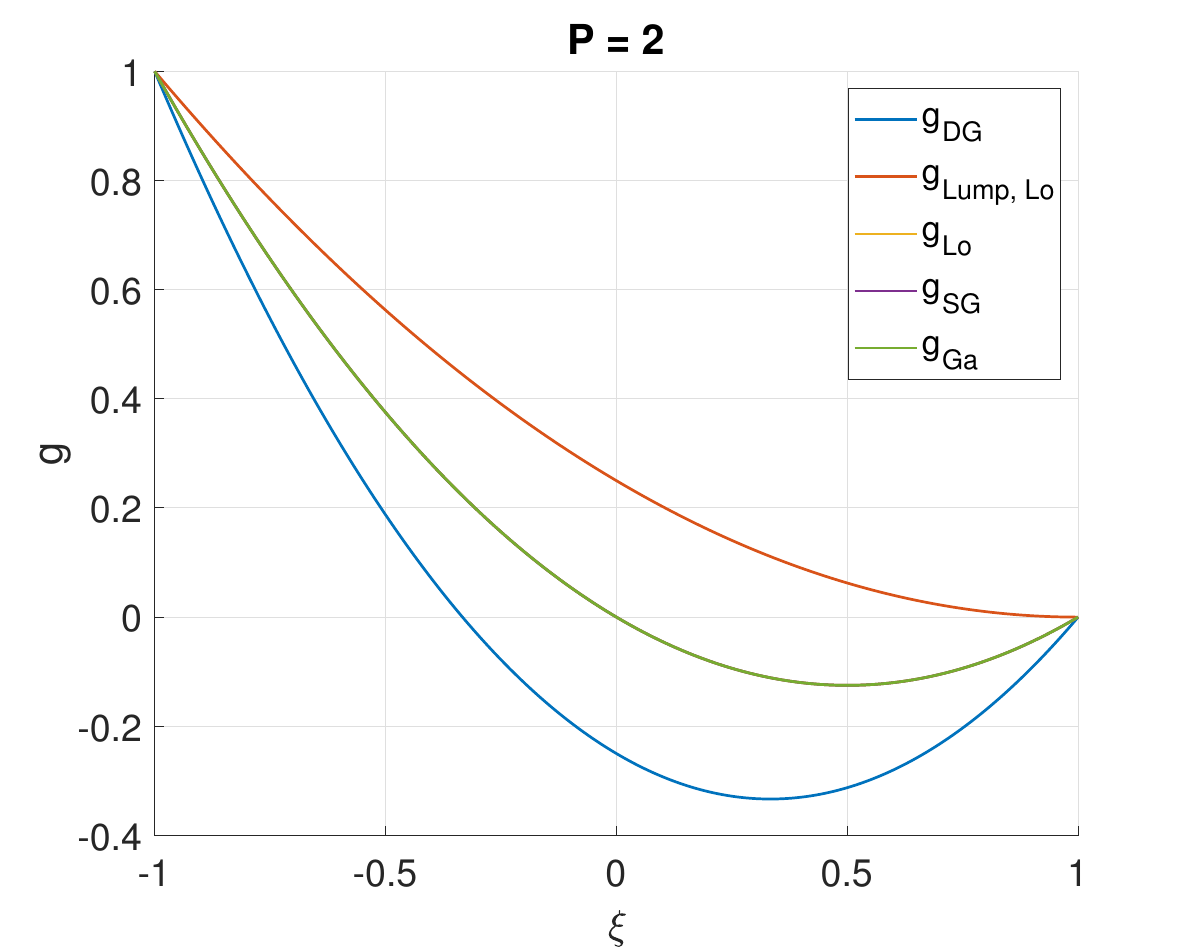}
    \end{minipage}\hfill
    \begin{minipage}{0.5\textwidth}
        \centering
        \includegraphics[width=\textwidth]{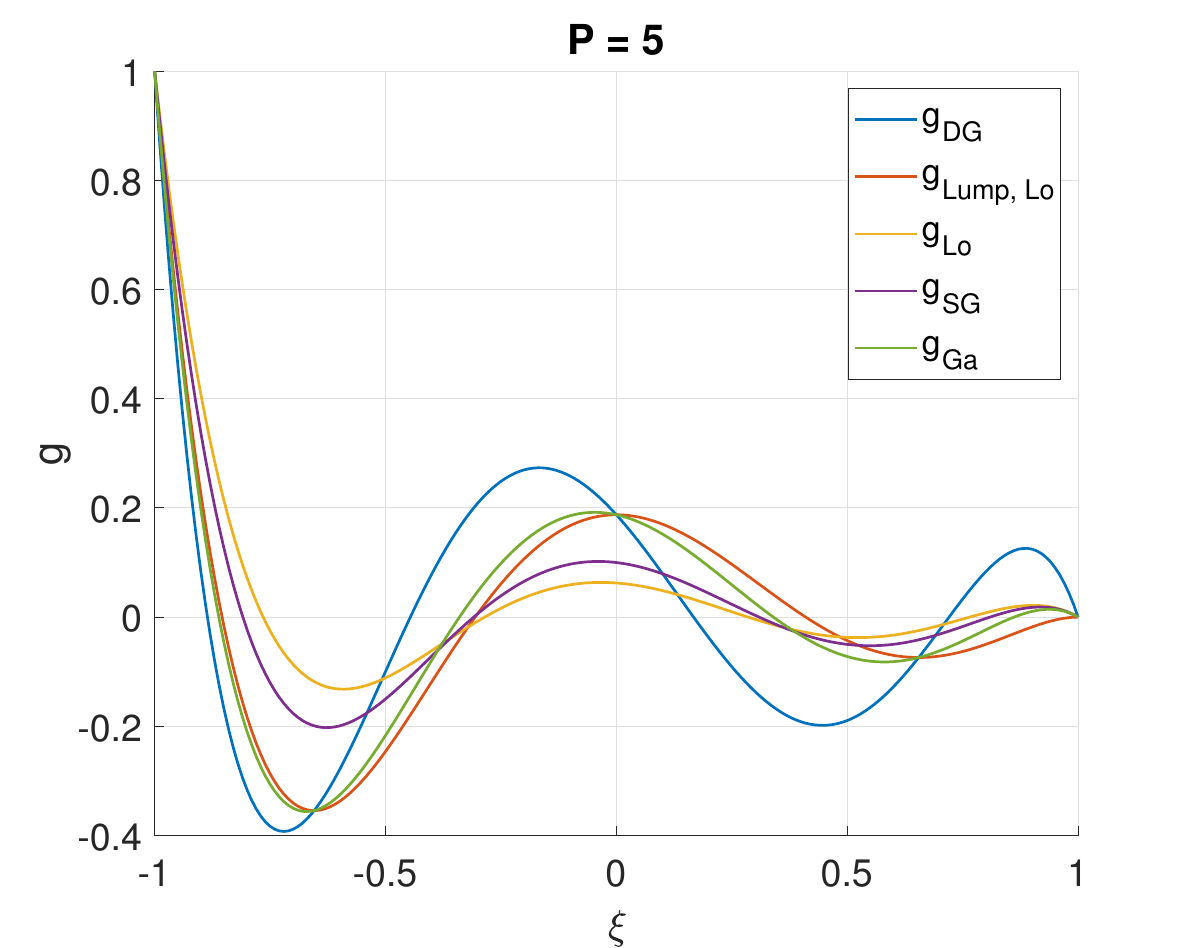}
    \end{minipage}
    \begin{minipage}{0.5\textwidth}
        \centering
        \includegraphics[width=\textwidth]{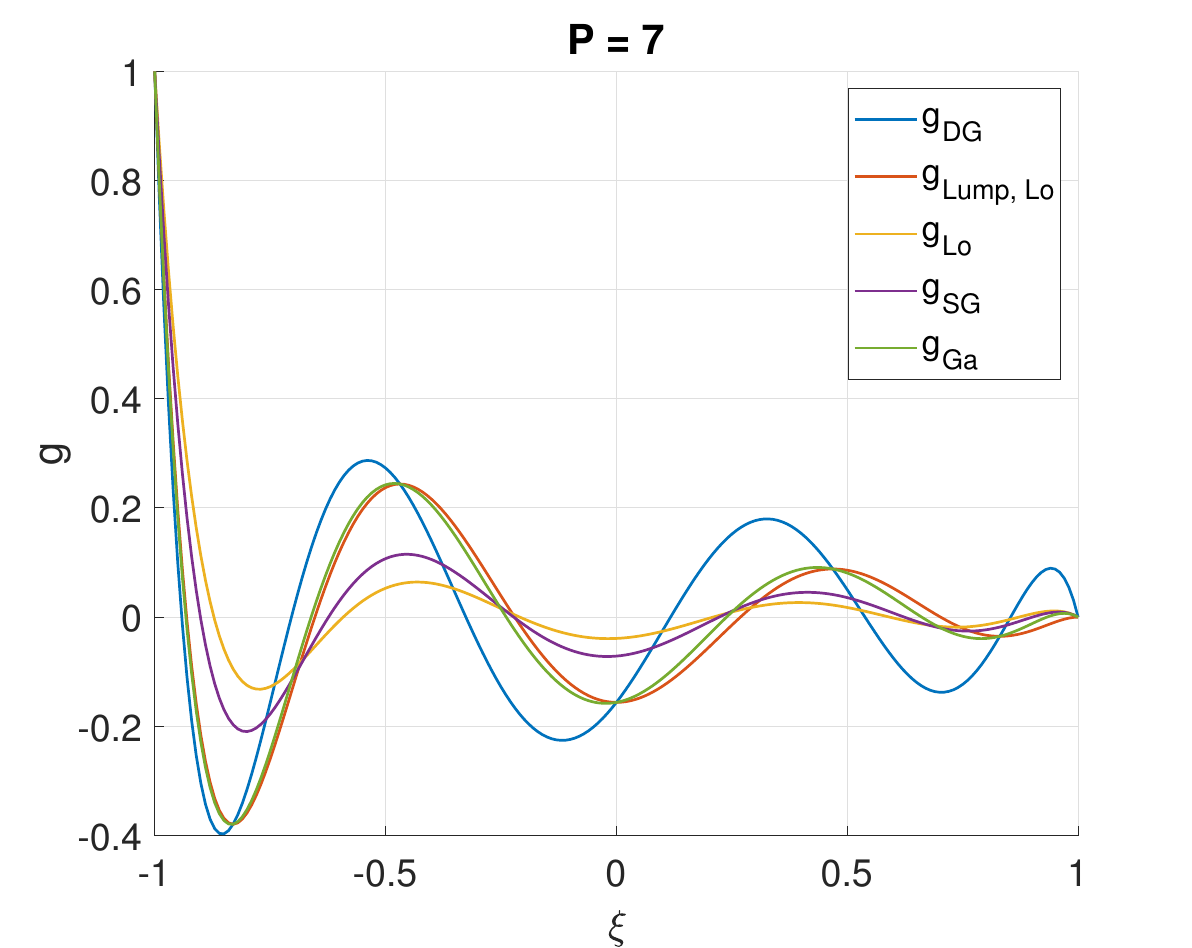}
    \end{minipage}\hfill
    \begin{minipage}{0.5\textwidth}
        \centering
        \includegraphics[width=\textwidth]{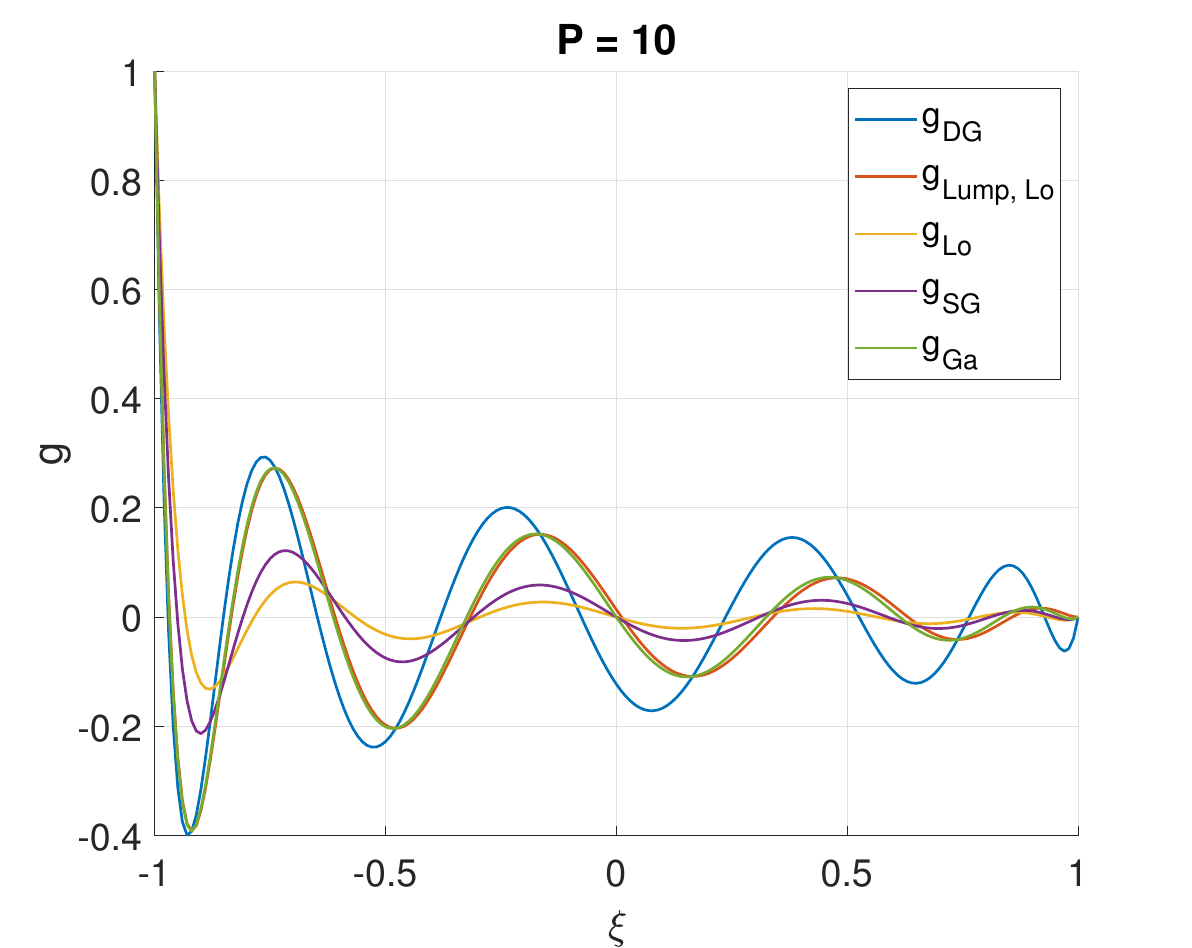}
    \end{minipage}
    \caption{Correction functions of multiple polynomial orders, $P$, for the interval $\xi \in [-1,1]$. For $P=2$, $g_{Lo} = g_{SG} = g_{Ga}$.}
    \label{Fig:gPlot}
\end{figure}

Plots of the correction functions of polynomial order, $P$, equal to $2$, $5$, $7$ and $10$ are shown in Fig.~\ref{Fig:gPlot}. For the second-order correction functions, $g_{Lo} = g_{SG} = g_{Ga}$. This means that, effectively, if only a linear interpolation is used for reconstructing the internal solution in a cell, then the usage of these three correction functions essentially result in the same scheme. Although not directly shown here, this is not the case anymore for $P \geq 3$.

From Eq.~(\ref{Eq:FullOperatorForm}), it is known that the stability domain of each scheme differ from each other according to the derivatives of the correction functions evaluated at the discrete nodes, at least in the context of linear stability analysis. In particular, observe that the highest gradients of the correction functions are achieved at the non-zero boundary. If this point is taken as a reference for comparing the derivatives of each correction function, then it is possible to say that, in general $g_{DG}' > g_{Ga}' > g_{Lump, Lo}' > g_{SG}' > g_{Lo}'$, where $g'$ denotes the derivative of $g$ in relation to $\xi$. As will be seen later, this is exactly the inverse relation that exists for the CFL limit of each scheme. In other words, the steeper the correction function is at the non-zero boundary, the lower the CFL limit is. Also notice that, as the order of the polynomials increases, the correction functions become even steeper, pointing towards the tendency of a reduction in the CFL limit as the order of the scheme increases.

\begin{figure}[h!]
    \centering
    \begin{minipage}{0.5\textwidth}
        \centering
        \includegraphics[width=\textwidth]{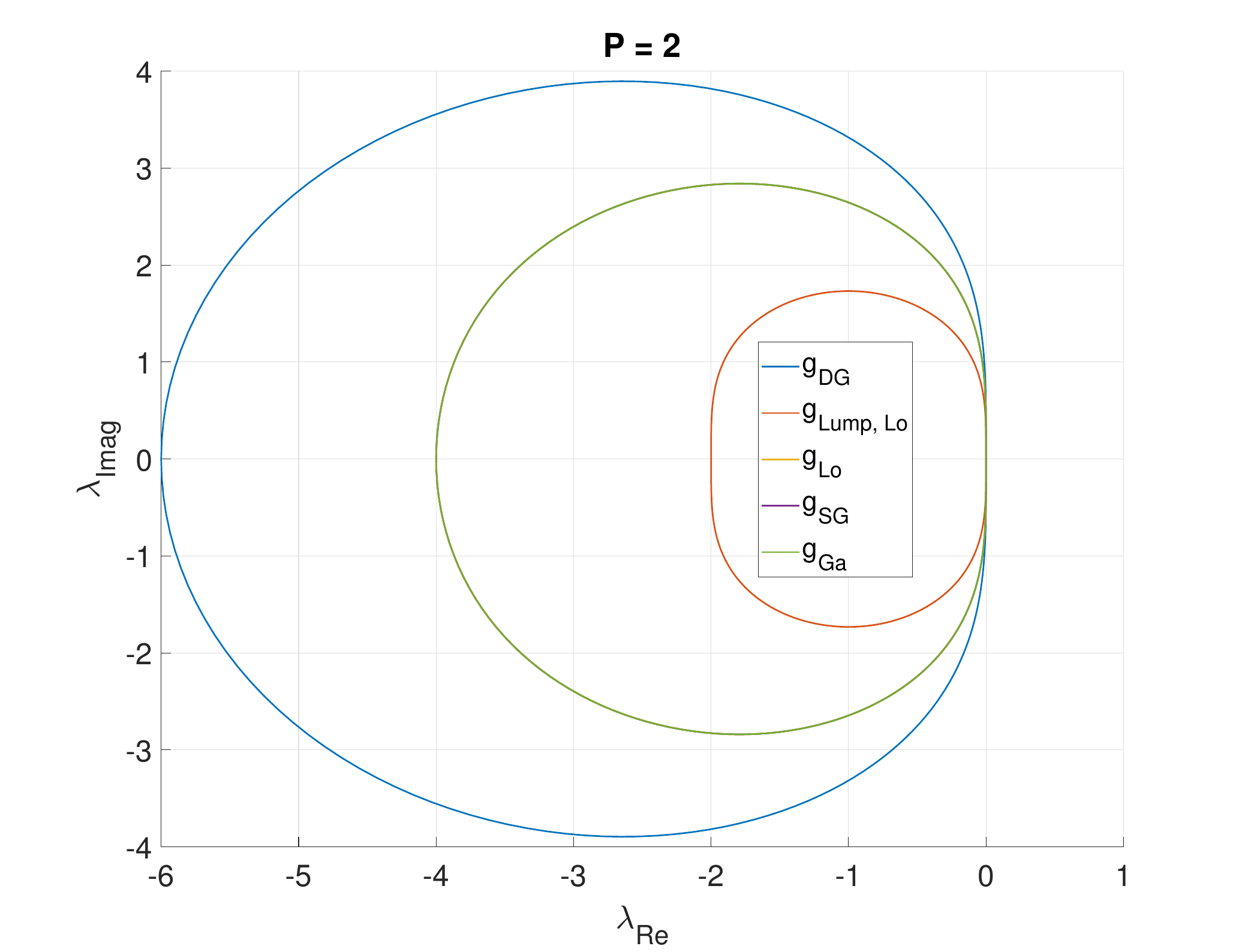}
        \label{Fig:eigP2}
    \end{minipage}\hfill
    \begin{minipage}{0.5\textwidth}
        \centering
        \includegraphics[width=\textwidth]{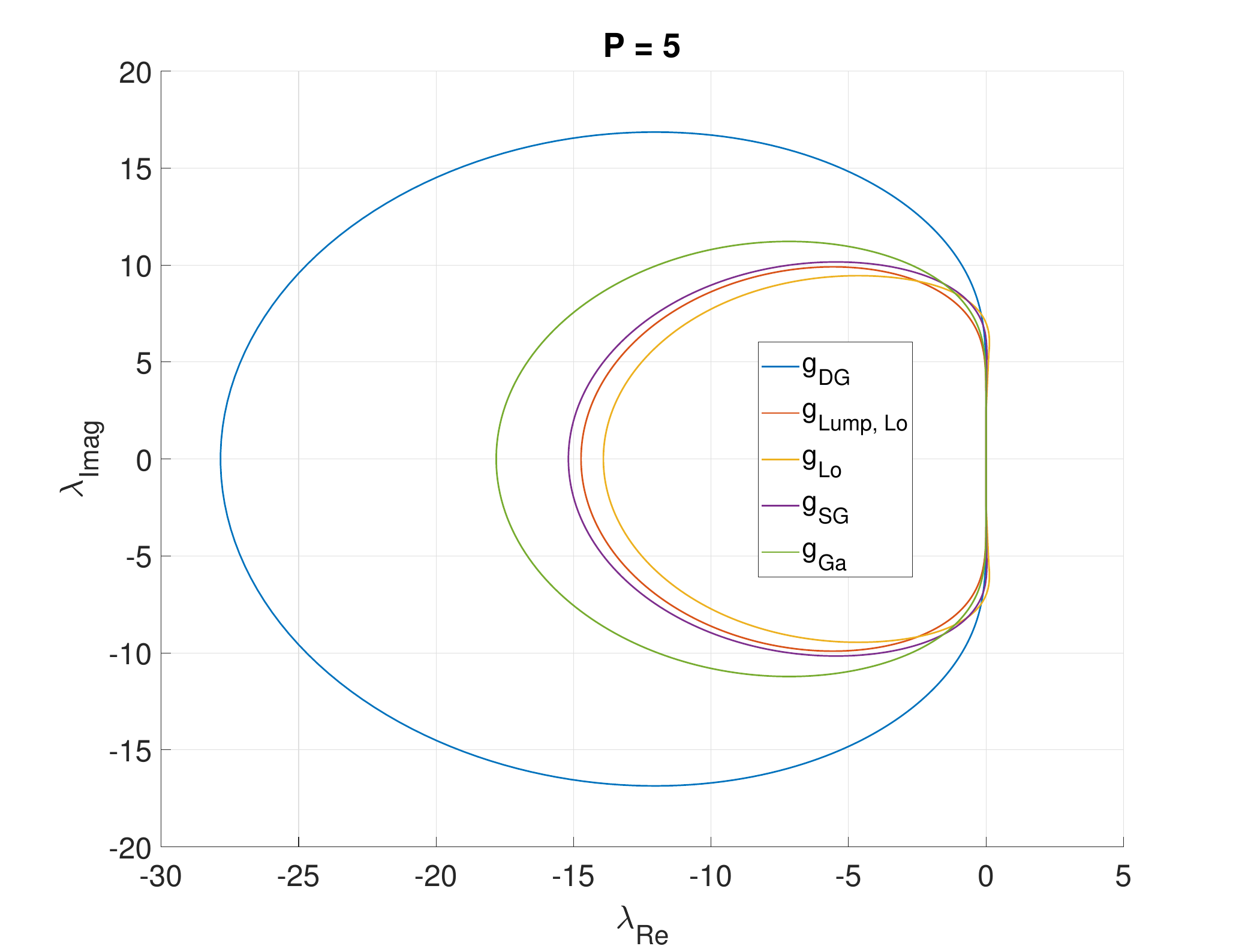}
        \label{Fig:eigP5}
    \end{minipage}
    \begin{minipage}{0.5\textwidth}
        \centering
        \includegraphics[width=\textwidth]{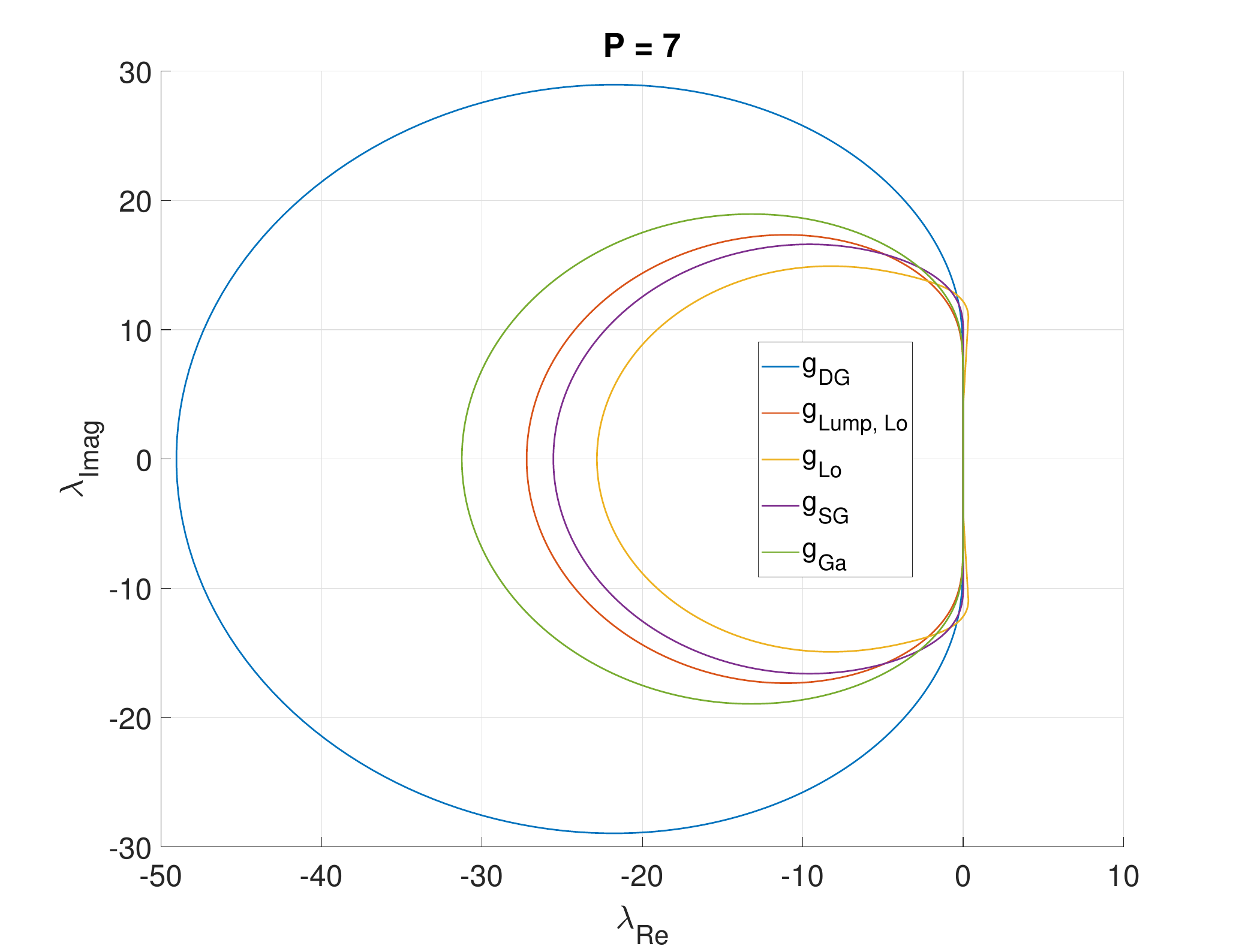}
        \label{Fig:eigP7}
    \end{minipage}\hfill
    \begin{minipage}{0.5\textwidth}
        \centering
        \includegraphics[width=\textwidth]{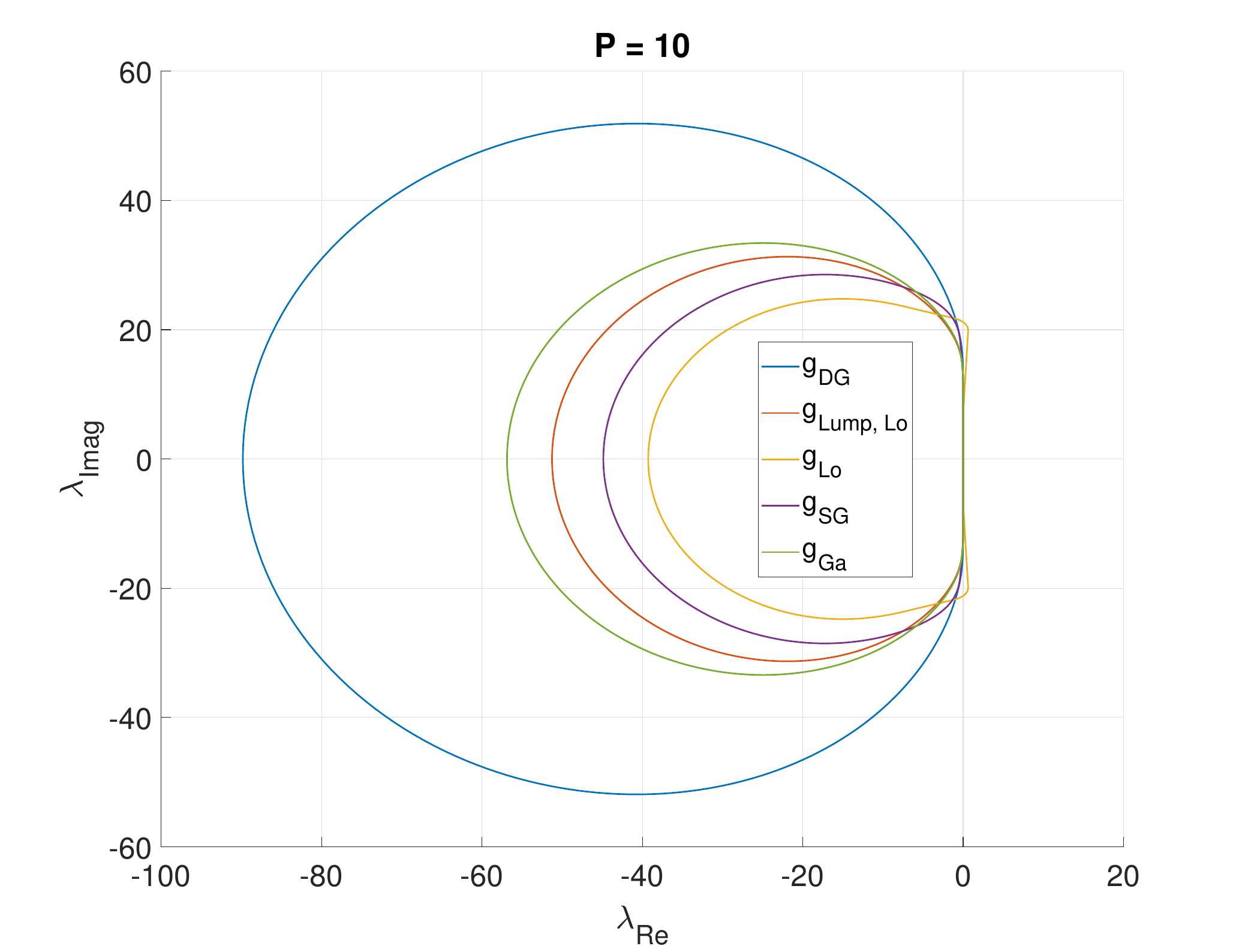}
        \label{Fig:eigP10}
    \end{minipage}
    \caption{Spectra of the principal eigenvalue, $\lambda_1$, of multiple polynomial orders, $P$, with $\omega \in [0, 2\pi]$. For $P=2$, the results are identical for $g_{Lo}$, $g_{SG}$ and $g_{Ga}$.}
    \label{Fig:eigPlot}
\end{figure}

Next, the spectra of the principal eigenvalue is plotted in Fig.~\ref{Fig:eigPlot} for correction functions of polynomial order equal to $2$, $5$, $7$ and $10$. For the advection equation, the order of the correction function coincides with the number of internal nodes of each cell, $P=K$. One of the properties that is apparent from this figure is that the region in which the principal eigenvalue is purely imaginary increases as the order of the correction polynomial also increases. From the previous section, it is known that the analytical solution to this problem is $\lambda_1 = -i \omega$, which is a purely imaginary number. This observation shows that, as expected, the capacity of each scheme to correctly represent a wider range of wave-numbers increases with the number of internal nodes per cell. However, for schemes constructed by using $g_{Lo}$ and $g_{SG}$, a mild instability can be observed as the order of the scheme increases. In the principal eigenvalue spectra, this instability manifests by the existence of positive, real eigenvalue components for a certain range of $\omega$ values.

In order to improve the data readability, the maximum real principal eigenvalue component of each scheme using $K$ internal node points per cell is shown in Tab.~\ref{Tab:MaxEig}. Notice that, for $K=3$ and above, schemes constructed with $g_{Lo}$ and $g_{SG}$ are effectively unstable. However, since the magnitude of these values is small even for the highest amount of internal node points considered here ($K=10$), in practice these schemes might be used, since by the time the unstable modes become effectively problematic, they would be already transported to the outside region of the computational domain. If the domain is effectively infinite, such as when a periodic boundary condition is used, then these schemes can also be used if the time-interval in which the time-march is performed is sufficiently small. If the time-march is carried indefinitely, then at some point the unstable modes will achieve a considerable magnitude, thus contaminating the results and making the solution diverge.

Another property that can be observed from the principal eigenvalue spectra is its minimum real value. It is clear that, in this regard, the $DG$ scheme achieves considerably lower values of real $\lambda_1$ components in relation to the other methods. Table \ref{Tab:MinEig} shows these values for $K$ varying from $2$ to $10$. As it can be seen, the relative order of magnitude of the correction function derivatives directly impacts the minimum value of the real component of the principal eigenvalue. That is, $Re \left( \lambda_{1_{DG}} \right) < Re \left( \lambda_{1_{Ga}} \right) < Re \left( \lambda_{1_{Lump, Lo}} \right) < Re \left( \lambda_{1_{SG}} \right) < Re \left( \lambda_{1_{Lo}} \right)$. This property directly affects the value of the CFL limit for each scheme. Table \ref{Tab:CFLLimits} shows the computed CFL Limits, according to linear stability analysis, for the advection equation when using different multistage Runge-Kutta schemes, going from the two-stage RK2 to the six-stage RK6, for $K$ going from $2$ to $10$. It is important to emphasize that, although a CFL limit has been defined for the schemes constructed with $g_{Lo}$ and $g_{SG}$, there are, as discussed before, certain wave-number that can excite unstable modes on these schemes, even if the CFL value is below the computed limit.

From Tab.~\ref{Tab:CFLLimits}, it is clear that the higher precision of the $DG$ scheme comes with a severe limitation in the maximum allowed CFL value. This can be a problem particularly if the high-order scheme is being used to solve an equation that has a stationary solution, since the usage of a lower CFL value usually, but not always, implies a lower convergence rate for the algorithm. In those cases, the usage of $g_{Lump, Lo}$ or $g_{Ga}$ might be more appropriate. If the transient solution is of interest, then the lower CFL limit of $g_{DG}$ might not be a problem, since other solution properties might end up limiting the maximum allowed CFL number to an even lower value.

The results from the analysis of order of accuracy of each scheme using Eq.~(\ref{Eq:ApproxAccuracy}) are shown in Tab.~(\ref{Tab:OrdAcc}). The values of $\omega$ used to approximately compute the scheme order are also shown in the same table. The phenomena of superconvergence \cite{roe2017} is readily observed from the given data, which enables the schemes to achieve an order that is much higher than $K$.  The DG scheme is clearly the one capable of achieving the highest order of spatial discretization error for a given $K$, followed by $g_{Lump, Lo}$ and $g_{Ga}$, as expected. For $K \geq 9$, the error terms go to machine zero for a wide range of ``small'' $\omega$ values when operating with 64 bit floating point numbers. Therefore, higher than adequate values of $\omega$ are used, which produces results that are not reliable and, therefore, are marked with a question mark (``?'').

\begin{figure}[h!]
\centering
\includegraphics[angle=0, width=10cm]{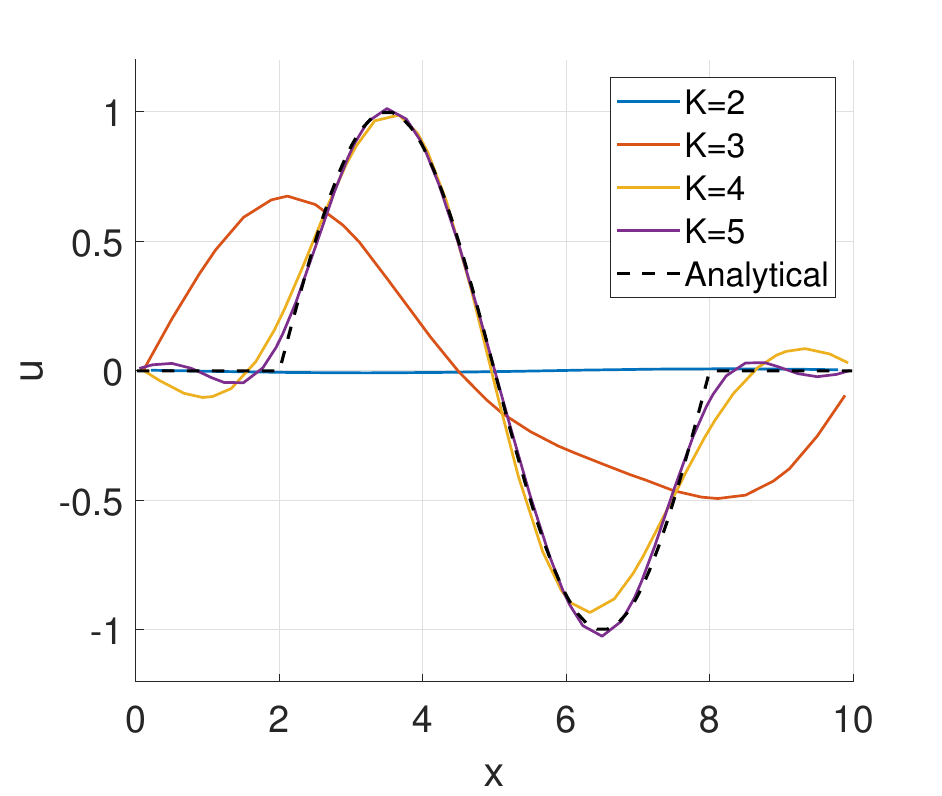}
\caption{Advection of a sine wave after $100$ flow through times using the correction function $g_{Ga}$.}
\label{Fig:SineWaveNumExp}
\end{figure}

Lastly, a numerical experiment is conducted in order to appreciate the effects of artificial dissipation that are acting over the discrete solution. In order to do that, a domain $x \in [0, 10]$ is discretized using $10$ cells of unitary length. The correction function is chosen to be $g_{Ga}$ and the boundary conditions of the problem are set to periodic. The initial condition is set to a sine wave of the form

\begin{equation}
\begin{cases}
u(x,0) = \sin \left( \pi \frac{\left( \xi - 2 \right)}{3} \right) \; \; \; \text{if} \; \; \; 2 \leq \xi \leq 8, \\
u(x,0) = 0, \; \; \; \text{otherwise.}
\end{cases},
\end{equation}

\noindent
the CFL value is set to $0.02$ for all cases, and the simulation is conducted for $100$ flow through times with $a=1$ using the RK4 time-march scheme. The results for values of $K$ from $2$ to $5$ are plotted in Fig.~(\ref{Fig:SineWaveNumExp}). As shown, the results related to the second-order scheme ($K=2$), are completely dissipated, with the solution being approximately flat when the simulation ends. For the forth-order scheme ($K=3$), the solution is considerably dissipated, which also resulted in the introduction of a significant phase error to it. For $K=3$, the value of the lowest point of the sine wave is approximately $50.56\%$ higher than the one defined by the analytical solution. For the sixth-order scheme ($K=4$), this difference is reduced to approximately $6.44\%$, while for the eighth-order scheme ($K=5$) this difference is reduced even further to $2.74\%$. Although not shown here, if $K=10$ was used, then this error would be only $0.03\%$. This shows that the resulting scheme does, indeed, introduce only extremely small amounts of artificial dissipation and is suited for applications where the conservation of a wave amplitude is desired.

\begin{table}[h!]
\centering
\caption{Maximum, real, eigenvalue component of each scheme using $K$ internal node points per cell. The schemes using $g_{Lo}$ and $g_{SG}$ are unstable when $K \geq 3$.}
\label{Tab:MaxEig}
\begin{tabular}{c|c|c|c|c|c}
\hline
              & \textbf{gDG} & \textbf{gLumpLo} & \textbf{gLo} & \textbf{gSG} & \textbf{gGa} \\ \hline
\textbf{K=2}  & 1.36E-30     & 0                & 1.87E-30     & 0            & 0            \\ \hline
\textbf{K=3}  & -7.97E-16    & 1.07E-15         & 0.008412     & 0.00283      & 3.68E-16     \\ \hline
\textbf{K=4}  & 1.54E-15     & -1.17E-15        & 0.049693     & 0.014644     & -4.17E-16    \\ \hline
\textbf{K=5}  & 4.00E-16     & 2.48E-15         & 0.122823     & 0.02813      & 4.53E-15     \\ \hline
\textbf{K=6}  & 1.82E-15     & 1.73E-15         & 0.218461     & 0.031044     & 1.38E-15     \\ \hline
\textbf{K=7}  & 6.16E-15     & 5.62E-15         & 0.326371     & 0.010877     & 3.27E-15     \\ \hline
\textbf{K=8}  & 4.56E-15     & 6.64E-15         & 0.43591      & 4.05E-05     & 6.00E-15     \\ \hline
\textbf{K=9}  & 4.12E-15     & 3.58E-15         & 0.537869     & 0.000373     & 5.27E-15     \\ \hline
\textbf{K=10} & 3.75E-15     & 4.83E-15         & 0.625159     & 0.001551     & 3.41E-15     \\ \hline
\end{tabular}
\end{table}

\begin{table}[h!]
\centering
\caption{Minimum, real, eigenvalue component of each scheme using $K$ internal node points per cell.}
\label{Tab:MinEig}
\begin{tabular}{c|c|c|c|c|c}
\hline
              & \textbf{gDG} & \textbf{gLumpLo} & \textbf{gLo} & \textbf{gSG} & \textbf{gGa} \\ \hline
\textbf{K=2}  & -6.0000      & -2.0000          & -4.0000      & -4.0000      & -4.0000      \\ \hline
\textbf{K=3}  & -11.8407     & -5.4196          & -6.8225      & -7.0998      & -7.6597      \\ \hline
\textbf{K=4}  & -19.1569     & -9.6485          & -10.1286     & -10.8399     & -12.2952     \\ \hline
\textbf{K=5}  & -27.8388     & -14.7291         & -13.9174     & -15.1889     & -17.8143     \\ \hline
\textbf{K=6}  & -37.8247     & -20.5985         & -18.1601     & -20.1037     & -24.1505     \\ \hline
\textbf{K=7}  & -49.0471     & -27.2132         & -22.8306     & -25.5481     & -31.2531     \\ \hline
\textbf{K=8}  & -61.4815     & -34.5459         & -27.9166     & -31.5035     & -39.0941     \\ \hline
\textbf{K=9}  & -75.0732     & -42.5679         & -33.3986     & -37.9448     & -47.6406     \\ \hline
\textbf{K=10} & -89.8181     & -51.2638         & -39.2711     & -44.8629     & -56.8784     \\ \hline
\end{tabular}
\end{table}

\begin{sidewaystable}[]
\centering
\caption{CFL limits of each scheme as a function of the number of internal nodes, $K$, and the chosen time-march procedure.}
\label{Tab:CFLLimits}
\begin{tabular}{cccccclcccccc}
\cline{1-6} \cline{8-13}
\multicolumn{1}{c|}{\textbf{K=2}}  & \multicolumn{1}{c|}{\textbf{gDG}} & \multicolumn{1}{c|}{\textbf{gLumpLo}} & \multicolumn{1}{c|}{\textbf{gLo}} & \multicolumn{1}{c|}{\textbf{gSG}} & \textbf{gGa}                 &  & \multicolumn{1}{c|}{\textbf{K=7}}  & \multicolumn{1}{c|}{\textbf{gDG}} & \multicolumn{1}{c|}{\textbf{gLumpLo}} & \multicolumn{1}{c|}{\textbf{gLo}} & \multicolumn{1}{c|}{\textbf{gSG}} & \textbf{gGa}         \\ \cline{1-6} \cline{8-13} 
\multicolumn{1}{c|}{\textbf{RK-2}} & \multicolumn{1}{c|}{0.333333}     & \multicolumn{1}{c|}{1}                & \multicolumn{1}{c|}{0.5}          & \multicolumn{1}{c|}{0.5}          & 0.5                          &  & \multicolumn{1}{c|}{\textbf{RK-2}} & \multicolumn{1}{c|}{0.040777}     & \multicolumn{1}{c|}{0.073494}         & \multicolumn{1}{c|}{0.087602}     & \multicolumn{1}{c|}{0.078284}     & 0.063994             \\ \cline{1-6} \cline{8-13} 
\multicolumn{1}{c|}{\textbf{RK-3}} & \multicolumn{1}{c|}{0.418791}     & \multicolumn{1}{c|}{1.256373}         & \multicolumn{1}{c|}{0.628186}     & \multicolumn{1}{c|}{0.628186}     & 0.628186                     &  & \multicolumn{1}{c|}{\textbf{RK-3}} & \multicolumn{1}{c|}{0.051231}     & \multicolumn{1}{c|}{0.092336}         & \multicolumn{1}{c|}{0.110060}     & \multicolumn{1}{c|}{0.098354}     & 0.080400             \\ \cline{1-6} \cline{8-13} 
\multicolumn{1}{c|}{\textbf{RK-4}} & \multicolumn{1}{c|}{0.464216}     & \multicolumn{1}{c|}{1.392647}         & \multicolumn{1}{c|}{0.696323}     & \multicolumn{1}{c|}{0.696323}     & 0.696323                     &  & \multicolumn{1}{c|}{\textbf{RK-4}} & \multicolumn{1}{c|}{0.056788}     & \multicolumn{1}{c|}{0.102351}         & \multicolumn{1}{c|}{0.121998}     & \multicolumn{1}{c|}{0.109022}     & 0.089121             \\ \cline{1-6} \cline{8-13} 
\multicolumn{1}{c|}{\textbf{RK-5}} & \multicolumn{1}{r|}{0.536175}     & \multicolumn{1}{r|}{1.608524}         & \multicolumn{1}{r|}{0.804262}     & \multicolumn{1}{r|}{0.804262}     & \multicolumn{1}{r}{0.804262} &  & \multicolumn{1}{c|}{\textbf{RK-5}} & \multicolumn{1}{c|}{0.065591}     & \multicolumn{1}{c|}{0.118216}         & \multicolumn{1}{c|}{0.140909}     & \multicolumn{1}{c|}{0.125921}     & 0.102935             \\ \cline{1-6} \cline{8-13} 
\multicolumn{1}{c|}{\textbf{RK-6}} & \multicolumn{1}{c|}{0.592240}     & \multicolumn{1}{c|}{1.776721}         & \multicolumn{1}{c|}{0.888360}     & \multicolumn{1}{c|}{0.888360}     & 0.888360                     &  & \multicolumn{1}{c|}{\textbf{RK-6}} & \multicolumn{1}{c|}{0.072450}     & \multicolumn{1}{c|}{0.130578}         & \multicolumn{1}{c|}{0.155644}     & \multicolumn{1}{c|}{0.139088}     & 0.113699             \\ \cline{1-6} \cline{8-13} 
\multicolumn{1}{l}{}               & \multicolumn{1}{l}{}              & \multicolumn{1}{l}{}                  & \multicolumn{1}{l}{}              & \multicolumn{1}{l}{}              & \multicolumn{1}{l}{}         &  & \multicolumn{1}{l}{}               & \multicolumn{1}{l}{}              & \multicolumn{1}{l}{}                  & \multicolumn{1}{l}{}              & \multicolumn{1}{l}{}              & \multicolumn{1}{l}{} \\ \cline{1-6} \cline{8-13} 
\multicolumn{1}{c|}{\textbf{K=3}}  & \multicolumn{1}{c|}{\textbf{gDG}} & \multicolumn{1}{c|}{\textbf{gLumpLo}} & \multicolumn{1}{c|}{\textbf{gLo}} & \multicolumn{1}{c|}{\textbf{gSG}} & \textbf{gGa}                 &  & \multicolumn{1}{c|}{\textbf{K=8}}  & \multicolumn{1}{c|}{\textbf{gDG}} & \multicolumn{1}{c|}{\textbf{gLumpLo}} & \multicolumn{1}{c|}{\textbf{gLo}} & \multicolumn{1}{c|}{\textbf{gSG}} & \textbf{gGa}         \\ \cline{1-6} \cline{8-13} 
\multicolumn{1}{c|}{\textbf{RK-2}} & \multicolumn{1}{c|}{0.168909}     & \multicolumn{1}{c|}{0.369029}         & \multicolumn{1}{c|}{0.29315}      & \multicolumn{1}{c|}{0.2817}       & 0.261107                     &  & \multicolumn{1}{c|}{\textbf{RK-2}} & \multicolumn{1}{c|}{0.03253}      & \multicolumn{1}{c|}{0.057894}         & \multicolumn{1}{c|}{0.071642}     & \multicolumn{1}{c|}{0.063485}     & 0.051159             \\ \cline{1-6} \cline{8-13} 
\multicolumn{1}{c|}{\textbf{RK-3}} & \multicolumn{1}{c|}{0.212213}     & \multicolumn{1}{c|}{0.463638}         & \multicolumn{1}{c|}{0.368305}     & \multicolumn{1}{c|}{0.353920}     & 0.328048                     &  & \multicolumn{1}{c|}{\textbf{RK-3}} & \multicolumn{1}{c|}{0.040870}     & \multicolumn{1}{c|}{0.072736}         & \multicolumn{1}{c|}{0.090009}     & \multicolumn{1}{c|}{0.079761}     & 0.064274             \\ \cline{1-6} \cline{8-13} 
\multicolumn{1}{c|}{\textbf{RK-4}} & \multicolumn{1}{c|}{0.235230}     & \multicolumn{1}{c|}{0.513927}         & \multicolumn{1}{c|}{0.408254}     & \multicolumn{1}{c|}{0.392308}     & 0.363630                     &  & \multicolumn{1}{c|}{\textbf{RK-4}} & \multicolumn{1}{c|}{0.045303}     & \multicolumn{1}{c|}{0.080626}         & \multicolumn{1}{c|}{0.099772}     & \multicolumn{1}{c|}{0.088412}     & 0.071246             \\ \cline{1-6} \cline{8-13} 
\multicolumn{1}{c|}{\textbf{RK-5}} & \multicolumn{1}{r|}{0.271694}     & \multicolumn{1}{r|}{0.593592}         & \multicolumn{1}{r|}{0.471539}     & \multicolumn{1}{r|}{0.453121}     & \multicolumn{1}{r}{0.419997} &  & \multicolumn{1}{c|}{\textbf{RK-5}} & \multicolumn{1}{c|}{0.052325}     & \multicolumn{1}{c|}{0.093124}         & \multicolumn{1}{c|}{0.115238}     & \multicolumn{1}{c|}{0.102117}     & 0.082290             \\ \cline{1-6} \cline{8-13} 
\multicolumn{1}{c|}{\textbf{RK-6}} & \multicolumn{1}{c|}{0.300104}     & \multicolumn{1}{c|}{0.655661}         & \multicolumn{1}{c|}{0.520845}     & \multicolumn{1}{c|}{0.500502}     & 0.463915                     &  & \multicolumn{1}{c|}{\textbf{RK-6}} & \multicolumn{1}{c|}{0.057797}     & \multicolumn{1}{c|}{0.102861}         & \multicolumn{1}{c|}{0.127288}     & \multicolumn{1}{c|}{0.112795}     & 0.090895             \\ \cline{1-6} \cline{8-13} 
\multicolumn{1}{l}{}               & \multicolumn{1}{l}{}              & \multicolumn{1}{l}{}                  & \multicolumn{1}{l}{}              & \multicolumn{1}{l}{}              & \multicolumn{1}{l}{}         &  & \multicolumn{1}{l}{}               & \multicolumn{1}{l}{}              & \multicolumn{1}{l}{}                  & \multicolumn{1}{l}{}              & \multicolumn{1}{l}{}              & \multicolumn{1}{l}{} \\ \cline{1-6} \cline{8-13} 
\multicolumn{1}{c|}{\textbf{K=4}}  & \multicolumn{1}{c|}{\textbf{gDG}} & \multicolumn{1}{c|}{\textbf{gLumpLo}} & \multicolumn{1}{c|}{\textbf{gLo}} & \multicolumn{1}{c|}{\textbf{gSG}} & \textbf{gGa}                 &  & \multicolumn{1}{c|}{\textbf{K=9}}  & \multicolumn{1}{c|}{\textbf{gDG}} & \multicolumn{1}{c|}{\textbf{gLumpLo}} & \multicolumn{1}{c|}{\textbf{gLo}} & \multicolumn{1}{c|}{\textbf{gSG}} & \textbf{gGa}         \\ \cline{1-6} \cline{8-13} 
\multicolumn{1}{c|}{\textbf{RK-2}} & \multicolumn{1}{c|}{0.104401}     & \multicolumn{1}{c|}{0.207286}         & \multicolumn{1}{c|}{0.197461}     & \multicolumn{1}{c|}{0.184504}     & 0.162665                     &  & \multicolumn{1}{c|}{\textbf{RK-2}} & \multicolumn{1}{c|}{0.026641}     & \multicolumn{1}{c|}{0.046984}         & \multicolumn{1}{c|}{0.059883}     & \multicolumn{1}{c|}{0.052708}     & 0.041981             \\ \cline{1-6} \cline{8-13} 
\multicolumn{1}{c|}{\textbf{RK-3}} & \multicolumn{1}{c|}{0.131167}     & \multicolumn{1}{c|}{0.260429}         & \multicolumn{1}{c|}{0.248084}     & \multicolumn{1}{c|}{0.231805}     & 0.204368                     &  & \multicolumn{1}{c|}{\textbf{RK-3}} & \multicolumn{1}{c|}{0.033471}     & \multicolumn{1}{c|}{0.059029}         & \multicolumn{1}{c|}{0.075235}     & \multicolumn{1}{c|}{0.066221}     & 0.052744             \\ \cline{1-6} \cline{8-13} 
\multicolumn{1}{c|}{\textbf{RK-4}} & \multicolumn{1}{c|}{0.145394}     & \multicolumn{1}{c|}{0.288676}         & \multicolumn{1}{c|}{0.274993}     & \multicolumn{1}{c|}{0.256948}     & 0.226535                     &  & \multicolumn{1}{c|}{\textbf{RK-4}} & \multicolumn{1}{c|}{0.037101}     & \multicolumn{1}{c|}{0.065432}         & \multicolumn{1}{c|}{0.083396}     & \multicolumn{1}{c|}{0.073404}     & 0.058465             \\ \cline{1-6} \cline{8-13} 
\multicolumn{1}{c|}{\textbf{RK-5}} & \multicolumn{1}{c|}{0.167932}     & \multicolumn{1}{c|}{0.333425}         & \multicolumn{1}{c|}{0.317620}     & \multicolumn{1}{c|}{0.296778}     & 0.261651                     &  & \multicolumn{1}{c|}{\textbf{RK-5}} & \multicolumn{1}{c|}{0.042852}     & \multicolumn{1}{c|}{0.075575}         & \multicolumn{1}{c|}{0.096323}     & \multicolumn{1}{c|}{0.084782}     & 0.067527             \\ \cline{1-6} \cline{8-13} 
\multicolumn{1}{c|}{\textbf{RK-6}} & \multicolumn{1}{c|}{0.185491}     & \multicolumn{1}{c|}{0.368290}         & \multicolumn{1}{c|}{0.350832}     & \multicolumn{1}{c|}{0.327811}     & 0.289010                     &  & \multicolumn{1}{c|}{\textbf{RK-6}} & \multicolumn{1}{c|}{0.047333}     & \multicolumn{1}{c|}{0.083477}         & \multicolumn{1}{c|}{0.106395}     & \multicolumn{1}{c|}{0.093648}     & 0.074589             \\ \cline{1-6} \cline{8-13} 
\multicolumn{1}{l}{}               & \multicolumn{1}{l}{}              & \multicolumn{1}{l}{}                  & \multicolumn{1}{l}{}              & \multicolumn{1}{l}{}              & \multicolumn{1}{l}{}         &  & \multicolumn{1}{l}{}               & \multicolumn{1}{l}{}              & \multicolumn{1}{l}{}                  & \multicolumn{1}{l}{}              & \multicolumn{1}{l}{}              & \multicolumn{1}{l}{} \\ \cline{1-6} \cline{8-13} 
\multicolumn{1}{c|}{\textbf{K=5}}  & \multicolumn{1}{c|}{\textbf{gDG}} & \multicolumn{1}{c|}{\textbf{gLumpLo}} & \multicolumn{1}{c|}{\textbf{gLo}} & \multicolumn{1}{c|}{\textbf{gSG}} & \textbf{gGa}                 &  & \multicolumn{1}{c|}{\textbf{K=10}} & \multicolumn{1}{c|}{\textbf{gDG}} & \multicolumn{1}{c|}{\textbf{gLumpLo}} & \multicolumn{1}{c|}{\textbf{gLo}} & \multicolumn{1}{c|}{\textbf{gSG}} & \textbf{gGa}         \\ \cline{1-6} \cline{8-13} 
\multicolumn{1}{c|}{\textbf{RK-2}} & \multicolumn{1}{c|}{0.071842}     & \multicolumn{1}{c|}{0.135786}         & \multicolumn{1}{c|}{0.143705}     & \multicolumn{1}{c|}{0.131675}     & 0.112269                     &  & \multicolumn{1}{c|}{\textbf{RK-2}} & \multicolumn{1}{c|}{0.022267}     & \multicolumn{1}{c|}{0.039014}         & \multicolumn{1}{c|}{0.050928}     & \multicolumn{1}{c|}{0.04458}      & 0.035163             \\ \cline{1-6} \cline{8-13} 
\multicolumn{1}{c|}{\textbf{RK-3}} & \multicolumn{1}{c|}{0.090261}     & \multicolumn{1}{c|}{0.170597}         & \multicolumn{1}{c|}{0.180547}     & \multicolumn{1}{c|}{0.165433}     & 0.141052                     &  & \multicolumn{1}{c|}{\textbf{RK-3}} & \multicolumn{1}{c|}{0.027976}     & \multicolumn{1}{c|}{0.049016}         & \multicolumn{1}{c|}{0.063985}     & \multicolumn{1}{c|}{0.056009}     & 0.044177             \\ \cline{1-6} \cline{8-13} 
\multicolumn{1}{c|}{\textbf{RK-4}} & \multicolumn{1}{c|}{0.100051}     & \multicolumn{1}{c|}{0.189101}         & \multicolumn{1}{c|}{0.200130}     & \multicolumn{1}{c|}{0.183377}     & 0.156352                     &  & \multicolumn{1}{c|}{\textbf{RK-4}} & \multicolumn{1}{c|}{0.031010}     & \multicolumn{1}{c|}{0.054333}         & \multicolumn{1}{c|}{0.070925}     & \multicolumn{1}{c|}{0.062085}     & 0.048969             \\ \cline{1-6} \cline{8-13} 
\multicolumn{1}{c|}{\textbf{RK-5}} & \multicolumn{1}{c|}{0.115560}     & \multicolumn{1}{c|}{0.218414}         & \multicolumn{1}{c|}{0.231153}     & \multicolumn{1}{c|}{0.211803}     & 0.180588                     &  & \multicolumn{1}{c|}{\textbf{RK-5}} & \multicolumn{1}{c|}{0.035817}     & \multicolumn{1}{c|}{0.062755}         & \multicolumn{1}{c|}{0.081919}     & \multicolumn{1}{c|}{0.071708}     & 0.056560             \\ \cline{1-6} \cline{8-13} 
\multicolumn{1}{c|}{\textbf{RK-6}} & \multicolumn{1}{c|}{0.127643}     & \multicolumn{1}{c|}{0.241253}         & \multicolumn{1}{c|}{0.255324}     & \multicolumn{1}{c|}{0.233950}     & 0.199471                     &  & \multicolumn{1}{c|}{\textbf{RK-6}} & \multicolumn{1}{c|}{0.039563}     & \multicolumn{1}{c|}{0.069317}         & \multicolumn{1}{c|}{0.090485}     & \multicolumn{1}{c|}{0.079207}     & 0.062474             \\ \cline{1-6} \cline{8-13} 
\multicolumn{1}{l}{}               & \multicolumn{1}{l}{}              & \multicolumn{1}{l}{}                  & \multicolumn{1}{l}{}              & \multicolumn{1}{l}{}              & \multicolumn{1}{l}{}         &  & \multicolumn{1}{l}{}               & \multicolumn{1}{l}{}              & \multicolumn{1}{l}{}                  & \multicolumn{1}{l}{}              & \multicolumn{1}{l}{}              & \multicolumn{1}{l}{} \\ \cline{1-6}
\multicolumn{1}{c|}{\textbf{K=6}}  & \multicolumn{1}{c|}{\textbf{gDG}} & \multicolumn{1}{c|}{\textbf{gLumpLo}} & \multicolumn{1}{c|}{\textbf{gLo}} & \multicolumn{1}{c|}{\textbf{gSG}} & \textbf{gGa}                 &  &                                    &                                   &                                       &                                   &                                   &                      \\ \cline{1-6}
\multicolumn{1}{c|}{\textbf{RK-2}} & \multicolumn{1}{c|}{0.052876}     & \multicolumn{1}{c|}{0.097094}         & \multicolumn{1}{c|}{0.110132}     & \multicolumn{1}{c|}{0.099484}     & 0.082814                     &  &                                    &                                   &                                       &                                   &                                   &                      \\ \cline{1-6}
\multicolumn{1}{c|}{\textbf{RK-3}} & \multicolumn{1}{c|}{0.066431}     & \multicolumn{1}{c|}{0.121987}         & \multicolumn{1}{c|}{0.138366}     & \multicolumn{1}{c|}{0.124989}     & 0.104045                     &  &                                    &                                   &                                       &                                   &                                   &                      \\ \cline{1-6}
\multicolumn{1}{c|}{\textbf{RK-4}} & \multicolumn{1}{c|}{0.073637}     & \multicolumn{1}{c|}{0.135218}         & \multicolumn{1}{c|}{0.153374}     & \multicolumn{1}{c|}{0.138546}     & 0.115331                     &  &                                    &                                   &                                       &                                   &                                   &                      \\ \cline{1-6}
\multicolumn{1}{c|}{\textbf{RK-5}} & \multicolumn{1}{c|}{0.085052}     & \multicolumn{1}{c|}{0.156179}         & \multicolumn{1}{c|}{0.177149}     & \multicolumn{1}{c|}{0.160023}     & 0.133208                     &  &                                    &                                   &                                       &                                   &                                   &                      \\ \cline{1-6}
\multicolumn{1}{c|}{\textbf{RK-6}} & \multicolumn{1}{c|}{0.093945}     & \multicolumn{1}{c|}{0.172510}         & \multicolumn{1}{c|}{0.195673}     & \multicolumn{1}{c|}{0.176756}     & 0.147137                     &  &                                    &                                   &                                       &                                   &                                   &                      \\ \cline{1-6}
\end{tabular}
\end{sidewaystable}

\begin{sidewaystable}[]
\small
\tabcolsep=0.11cm
\centering
\caption{Order of accuracy analysis of each scheme as a function of the number of internal nodes, $K$. }
\label{Tab:OrdAcc}
\begin{tabular}{cccccc}
\hline
\multicolumn{1}{c|}{\textbf{K=2}}                      & \multicolumn{1}{c|}{\textbf{gDG}}                           & \multicolumn{1}{c|}{\textbf{gLumpLo}}                      & \multicolumn{1}{c|}{\textbf{gLo}}                          & \multicolumn{1}{c|}{\textbf{gSG}}                           & \textbf{gGa}                          \\ \hline
\multicolumn{1}{c|}{\textbf{Err. Coarse (w = 0.1*pi)}} & \multicolumn{1}{c|}{-0.000133848 + i*-1.10632e-05}          & \multicolumn{1}{c|}{-0.00109372 + i*0.00480392}            & \multicolumn{1}{c|}{-0.000294249 + i*0.00124472}           & \multicolumn{1}{c|}{-0.000294249 + i*0.00124472}            & -0.000294249 + i*0.00124472           \\ \hline
\multicolumn{1}{c|}{\textbf{Err. Fine (w = 0.05*pi)}}  & \multicolumn{1}{c|}{-8.43263e-06 + i*-3.52035e-07}          & \multicolumn{1}{c|}{-7.39806e-05 + i*0.000633594}          & \multicolumn{1}{c|}{-1.88609e-05 + i*0.000159965}          & \multicolumn{1}{c|}{-1.88609e-05 + i*0.000159965}           & -1.88609e-05 + i*0.000159965          \\ \hline
\multicolumn{1}{c|}{\textbf{Approx. Ord. Accurary}}    & \multicolumn{1}{c|}{3 (2.99212)}                            & \multicolumn{1}{c|}{2 (1.94927)}                           & \multicolumn{1}{c|}{2 (1.98926)}                           & \multicolumn{1}{c|}{2 (1.98926)}                            & 2 (1.98926)                           \\ \hline
\multicolumn{1}{l}{}                                   & \multicolumn{1}{l}{}                                        & \multicolumn{1}{l}{}                                       & \multicolumn{1}{l}{}                                       & \multicolumn{1}{l}{}                                        & \multicolumn{1}{l}{}                  \\ \hline
\multicolumn{1}{c|}{\textbf{K=3}}                      & \multicolumn{1}{c|}{\textbf{gDG}}                           & \multicolumn{1}{c|}{\textbf{gLumpLo}}                      & \multicolumn{1}{c|}{\textbf{gLo}}                          & \multicolumn{1}{c|}{\textbf{gSG}}                           & \textbf{gGa}                          \\ \hline
\multicolumn{1}{c|}{\textbf{Err. Coarse (w = 0.1*pi)}} & \multicolumn{1}{c|}{-1.32737e-07 + i*-7.16148e-09}          & \multicolumn{1}{c|}{-8.2924e-07 + i*6.29083e-06}           & \multicolumn{1}{c|}{2.64043e-05 + i*5.62676e-06}           & \multicolumn{1}{c|}{1.63614e-05 + i*4.56468e-06}            & -3.68845e-07 + i*2.80021e-06          \\ \hline
\multicolumn{1}{c|}{\textbf{Err. Fine (w = 0.05*pi)}}  & \multicolumn{1}{c|}{-2.08326e-09 + i*-5.61232e-11}          & \multicolumn{1}{c|}{-1.30194e-08 + i*1.98573e-07}          & \multicolumn{1}{c|}{1.68093e-06 + i*1.76785e-07}           & \multicolumn{1}{c|}{1.04836e-06 + i*1.43581e-07}            & -5.78744e-09 + i*8.82872e-08          \\ \hline
\multicolumn{1}{c|}{\textbf{Approx. Ord. Accurary}}    & \multicolumn{1}{c|}{5 (4.99516)}                            & \multicolumn{1}{c|}{4 (3.99485)}                           & \multicolumn{1}{c|}{3 (2.99755)}                           & \multicolumn{1}{c|}{3 (3.00476)}                            & 4 (3.9965)                            \\ \hline
\multicolumn{1}{l}{}                                   & \multicolumn{1}{l}{}                                        & \multicolumn{1}{l}{}                                       & \multicolumn{1}{l}{}                                       & \multicolumn{1}{l}{}                                        & \multicolumn{1}{l}{}                  \\ \hline
\multicolumn{1}{c|}{\textbf{K=4}}                      & \multicolumn{1}{c|}{\textbf{gDG}}                           & \multicolumn{1}{c|}{\textbf{gLumpLo}}                      & \multicolumn{1}{c|}{\textbf{gLo}}                          & \multicolumn{1}{c|}{\textbf{gSG}}                           & \textbf{gGa}                          \\ \hline
\multicolumn{1}{c|}{\textbf{Err. Coarse (w = 0.5*pi)}} & \multicolumn{1}{c|}{-2.37209e-05 + i*-4.84039e-06}          & \multicolumn{1}{c|}{-0.000127768 + i*0.000255549}          & \multicolumn{1}{c|}{0.00114085 + i*-0.000751305}           & \multicolumn{1}{c|}{0.000631333 + i*-0.000375493}           & -7.26229e-05 + i*0.000145976          \\ \hline
\multicolumn{1}{c|}{\textbf{Err. Fine (w = 0.25*pi)}}  & \multicolumn{1}{c|}{-1.00038e-07 + i*-1.00336e-08}          & \multicolumn{1}{c|}{-5.43425e-07 + i*2.32315e-06}          & \multicolumn{1}{c|}{2.07662e-05 + i*-3.88132e-05}          & \multicolumn{1}{c|}{1.19741e-05 + i*-2.20701e-05}           & -3.06404e-07 + i*1.31144e-06          \\ \hline
\multicolumn{1}{c|}{\textbf{Approx. Ord. Accurary}}    & \multicolumn{1}{c|}{7 (6.91168)}                            & \multicolumn{1}{c|}{6 (5.90389)}                           & \multicolumn{1}{c|}{4 (3.95569)}                           & \multicolumn{1}{c|}{4 (3.87059)}                            & 6 (5.91962)                           \\ \hline
\multicolumn{1}{l}{}                                   & \multicolumn{1}{l}{}                                        & \multicolumn{1}{l}{}                                       & \multicolumn{1}{l}{}                                       & \multicolumn{1}{l}{}                                        & \multicolumn{1}{l}{}                  \\ \hline
\multicolumn{1}{c|}{\textbf{K=5}}                      & \multicolumn{1}{c|}{\textbf{gDG}}                           & \multicolumn{1}{c|}{\textbf{gLumpLo}}                      & \multicolumn{1}{c|}{\textbf{gLo}}                          & \multicolumn{1}{c|}{\textbf{gSG}}                           & \textbf{gGa}                          \\ \hline
\multicolumn{1}{c|}{\textbf{Err. Coarse (w = 0.5*pi)}} & \multicolumn{1}{c|}{-1.85263e-07 + i*-2.98036e-08}          & \multicolumn{1}{c|}{-9.31827e-07 + i*2.50776e-06}          & \multicolumn{1}{c|}{-5.64685e-05 + i*-7.78105e-05}         & \multicolumn{1}{c|}{-2.6189e-05 + i*-4.03572e-05}           & -5.9944e-07 + i*1.6161e-06            \\ \hline
\multicolumn{1}{c|}{\textbf{Err. Fine (w = 0.25*pi)}}  & \multicolumn{1}{c|}{-1.91636e-10 + i*-1.52549e-11}          & \multicolumn{1}{c|}{-9.68607e-10 + i*5.41507e-09}          & \multicolumn{1}{c|}{-1.37175e-06 + i*-6.991e-07}           & \multicolumn{1}{c|}{-6.67361e-07 + i*-3.69313e-07}          & -6.20698e-10 + i*3.47148e-09          \\ \hline
\multicolumn{1}{c|}{\textbf{Approx. Ord. Accurary}}    & \multicolumn{1}{c|}{9 (8.93087)}                            & \multicolumn{1}{c|}{8 (7.92578)}                           & \multicolumn{1}{c|}{5 (4.96451)}                           & \multicolumn{1}{c|}{5 (4.97901)}                            & 8 (7.93304)                           \\ \hline
\multicolumn{1}{l}{}                                   & \multicolumn{1}{l}{}                                        & \multicolumn{1}{l}{}                                       & \multicolumn{1}{l}{}                                       & \multicolumn{1}{l}{}                                        & \multicolumn{1}{l}{}                  \\ \hline
\multicolumn{1}{c|}{\textbf{K=6}}                      & \multicolumn{1}{c|}{\textbf{gDG}}                           & \multicolumn{1}{c|}{\textbf{gLumpLo}}                      & \multicolumn{1}{c|}{\textbf{gLo}}                          & \multicolumn{1}{c|}{\textbf{gSG}}                           & \textbf{gGa}                          \\ \hline
\multicolumn{1}{c|}{\textbf{Err. Coarse (w = 0.5*pi)}} & \multicolumn{1}{c|}{-9.5894e-10 + i*-1.27578e-10}           & \multicolumn{1}{c|}{-4.61976e-09 + i*1.55294e-08}          & \multicolumn{1}{c|}{-4.56487e-06 + i*3.49946e-06}          & \multicolumn{1}{c|}{-2.09356e-06 + i*1.5715e-06}            & -3.21779e-09 + i*1.08248e-08          \\ \hline
\multicolumn{1}{c|}{\textbf{Err. Fine (w = 0.25*pi)}}  & \multicolumn{1}{c|}{-2.45897e-13 + i*-1.79856e-14}          & \multicolumn{1}{c|}{-1.18275e-12 + i*8.16502e-12}          & \multicolumn{1}{c|}{-2.02894e-08 + i*4.11367e-08}          & \multicolumn{1}{c|}{-9.38831e-09 + i*1.89655e-08}           & -8.23232e-13 + i*5.67724e-12          \\ \hline
\multicolumn{1}{c|}{\textbf{Approx. Ord. Accurary}}    & \multicolumn{1}{c|}{11 (10.938)}                            & \multicolumn{1}{c|}{10 (9.93945)}                          & \multicolumn{1}{c|}{6 (5.9704)}                            & \multicolumn{1}{c|}{6 (5.95071)}                            & 10 (9.94294)                          \\ \hline
\multicolumn{1}{l}{}                                   & \multicolumn{1}{l}{}                                        & \multicolumn{1}{l}{}                                       & \multicolumn{1}{l}{}                                       & \multicolumn{1}{l}{}                                        & \multicolumn{1}{l}{}                  \\ \hline
\multicolumn{1}{c|}{\textbf{K=7}}                      & \multicolumn{1}{c|}{\textbf{gDG}}                           & \multicolumn{1}{c|}{\textbf{gLumpLo}}                      & \multicolumn{1}{c|}{\textbf{gLo}}                          & \multicolumn{1}{c|}{\textbf{gSG}}                           & \textbf{gGa}                          \\ \hline
\multicolumn{1}{c|}{\textbf{Err. Coarse (w = 0.9*pi)}} & \multicolumn{1}{c|}{-1.18051e-08 + i*-2.44948e-09}          & \multicolumn{1}{c|}{-5.47977e-08 + i*1.15921e-07}          & \multicolumn{1}{c|}{-1.45745e-06 + i*3.12364e-05}          & \multicolumn{1}{c|}{-1.0471e-06 + i*1.32422e-05}            & -4.05099e-08 + i*8.5812e-08           \\ \hline
\multicolumn{1}{c|}{\textbf{Err. Fine (w = 0.45*pi)}}  & \multicolumn{1}{c|}{-8.14129e-13 + i*-8.4821e-14}           & \multicolumn{1}{c|}{-3.82377e-12 + i*1.71563e-11}          & \multicolumn{1}{c|}{8.93419e-08 + i*9.34842e-08}           & \multicolumn{1}{c|}{3.67039e-08 + i*4.0442e-08}             & -2.81076e-12 + i*1.26292e-11          \\ \hline
\multicolumn{1}{c|}{\textbf{Approx. Ord. Accurary}}    & \multicolumn{1}{c|}{13 (12.8464)}                           & \multicolumn{1}{c|}{12 (11.8326)}                          & \multicolumn{1}{c|}{7 (6.91781)}                           & \multicolumn{1}{c|}{7 (6.92615)}                            & 12 (11.8405)                          \\ \hline
\multicolumn{1}{l}{}                                   & \multicolumn{1}{l}{}                                        & \multicolumn{1}{l}{}                                       & \multicolumn{1}{l}{}                                       & \multicolumn{1}{l}{}                                        & \multicolumn{1}{l}{}                  \\ \hline
\multicolumn{1}{c|}{\textbf{K=8}}                      & \multicolumn{1}{c|}{\textbf{gDG}}                           & \multicolumn{1}{c|}{\textbf{gLumpLo}}                      & \multicolumn{1}{c|}{\textbf{gLo}}                          & \multicolumn{1}{c|}{\textbf{gSG}}                           & \textbf{gGa}                          \\ \hline
\multicolumn{1}{c|}{\textbf{Err. Coarse (w = 0.9*pi)}} & \multicolumn{1}{c|}{-1.07403e-10 + i*-1.93716e-11}          & \multicolumn{1}{c|}{-4.88962e-10 + i*1.22004e-09}          & \multicolumn{1}{c|}{2.60394e-06 + i*5.40995e-08}           & \multicolumn{1}{c|}{1.01699e-06 + i*4.3816e-08}             & -3.75909e-10 + i*9.38631e-10          \\ \hline
\multicolumn{1}{c|}{\textbf{Err. Fine (w = 0.45*pi)}}  & \multicolumn{1}{c|}{2.69351e-16 + i*-8.88178e-16}           & \multicolumn{1}{c|}{-9.7475e-15 + i*4.08562e-14}           & \multicolumn{1}{c|}{3.8169e-09 + i*-3.74164e-09}           & \multicolumn{1}{c|}{1.51576e-09 + i*-1.47441e-09}           & -5.79045e-15 + i*3.15303e-14          \\ \hline
\multicolumn{1}{c|}{\textbf{Approx. Ord. Accurary}}    & \multicolumn{1}{c|}{16 (15.8434)}                           & \multicolumn{1}{c|}{14 (13.9335)}                          & \multicolumn{1}{c|}{8 (7.92862)}                           & \multicolumn{1}{c|}{8 (7.91105)}                            & 14 (13.9449)                 \\ \hline
\multicolumn{1}{l}{\textbf{}}                          & \multicolumn{1}{l}{}                                        & \multicolumn{1}{l}{}                                       & \multicolumn{1}{l}{}                                       & \multicolumn{1}{l}{}                                        & \multicolumn{1}{l}{}                  \\ \hline
\multicolumn{1}{c|}{\textbf{K=9}}                      & \multicolumn{1}{c|}{\textbf{gDG}}                           & \multicolumn{1}{c|}{\textbf{gLumpLo}}                      & \multicolumn{1}{c|}{\textbf{gLo}}                          & \multicolumn{1}{c|}{\textbf{gSG}}                           & \textbf{gGa}                          \\ \hline
\multicolumn{1}{c|}{\textbf{Err. Coarse (w = 1.7*pi)}} & \multicolumn{1}{c|}{-5.11208e-08 + i*-1.59577e-08}          & \multicolumn{1}{c|}{-2.25192e-07 + i*3.01006e-07}          & \multicolumn{1}{c|}{9.5005e-05 + i*-1.61984e-05}           & \multicolumn{1}{c|}{3.49559e-05 + i*-4.82122e-06}           & -1.79825e-07 + i*2.40834e-07          \\ \hline
\multicolumn{1}{c|}{\textbf{Err. Fine (w = 0.85*pi)}}  & \multicolumn{1}{c|}{-2.77972e-13 + i*-4.35207e-14}          & \multicolumn{1}{c|}{-1.2318e-12 + i*3.75078e-12}           & \multicolumn{1}{c|}{-9.536e-09 + i*-1.09818e-07}           & \multicolumn{1}{c|}{-2.77138e-09 + i*-4.11502e-08}          & -9.74848e-13 + i*2.97495e-12          \\ \hline
\multicolumn{1}{c|}{\textbf{Approx. Ord. Accurary}}    & \multicolumn{1}{c|}{17 (16.5382) ?}                         & \multicolumn{1}{c|}{16 (15.539) ?}                         & \multicolumn{1}{c|}{9 (8.772) ?}                           & \multicolumn{1}{c|}{9 (8.74075) ?}                          & 16 (15.5509) ?                        \\ \hline
\multicolumn{1}{l}{\textbf{}}                          & \multicolumn{1}{l}{}                                        & \multicolumn{1}{l}{}                                       & \multicolumn{1}{l}{}                                       & \multicolumn{1}{l}{}                                        & \multicolumn{1}{l}{}                  \\ \hline
\multicolumn{1}{c|}{\textbf{K=10}}                     & \multicolumn{1}{c|}{\textbf{gDG}}                           & \multicolumn{1}{c|}{\textbf{gLumpLo}}                      & \multicolumn{1}{c|}{\textbf{gLo}}                          & \multicolumn{1}{c|}{\textbf{gSG}}                           & \textbf{gGa}                          \\ \hline
\multicolumn{1}{c|}{\textbf{Err. Coarse (w = 1.7*pi)}} & \multicolumn{1}{c|}{-1.06301e-09 + i*-2.95583e-10}          & \multicolumn{1}{c|}{-4.64451e-09 + i*7.21506e-09}          & \multicolumn{1}{c|}{-2.48094e-06 + i*-1.22152e-05}         & \multicolumn{1}{c|}{-7.37971e-07 + i*-4.32031e-06}          & -3.79089e-09 + i*5.8958e-09           \\ \hline
\multicolumn{1}{c|}{\textbf{Err. Fine (w = 0.85*pi)}}  & \multicolumn{1}{c|}{-1.69582e-15 + i*3.10862e-15}           & \multicolumn{1}{c|}{-9.81563e-15 + i*2.30926e-14}          & \multicolumn{1}{c|}{-6.97308e-09 + i*7.11174e-10}          & \multicolumn{1}{c|}{-2.45462e-09 + i*2.20544e-10}           & -1.02293e-14 + i*1.33227e-14          \\ \hline
\multicolumn{1}{c|}{\textbf{Approx. Ord. Accurary}}    & \multicolumn{1}{c|}{? (17.2493) ?}                          & \multicolumn{1}{c|}{? (17.3835) ?}                         & \multicolumn{1}{c|}{10 (9.79629) ?}                        & \multicolumn{1}{c|}{10 (9.79636) ?}                         & 18 (17.6707) ?                        \\ \hline
\end{tabular}
\end{sidewaystable}

\section{CONCLUDING REMARKS}

In the present work, five different schemes have been constructed using the Flux Reconstruction framework and evaluated using linear Fourier stability analysis for different values of internal node points per cell, $K$. It was observed that, the steeper the correction function used by each scheme is, the lower is the CFL limit associated with the resulting scheme. Out of all schemes tested, the $DG$ was the one that achieved the highest order of accuracy for a given $K$. This property, however, is achieved at the cost of severely reducing the maximum allowed CFL value for keeping the scheme stable. If only the stationary solution is required, then it might be better to use $g_{Lump, Lo}$ or $g_{Ga}$, since both of them allow the usage of higher CFL values without compromising too much on the order of accuracy of the scheme. Schemes constructed by using $g_{Lo}$ and $g_{SG}$ were seen to be mild unstable for $K \geq 3$, but can still be used in situations where the unstable modes are unable to develop. Values for the CFL limit as well as orders of accuracy of each scheme have been computed and tabulated, the former being of particular interest in the execution of multi-dimensional, high-fidelity simulations that employ high-order schemes, since it can be used as a first estimation for the CFL value to be employed. Finally, the effects of artificial dissipation have been observed in a simple numerical experiment, where the second-order scheme completely destroyed the transport of a wave before the simulation finished.

In the current context, future work will address similar analysis in a space of higher dimensions. Furthermore, investigations can be conducted using either the Euler equations or the Navier-Stokes equations in order to establish how well the current results can be extrapolated to those systems of equations.

\section*{ACKNOWLEDGEMENTS}
The authors wish to express their gratitude to Funda\c{c}\~{a}o de Amparo \`{a} Pesquisa do Estado de 
S\~{a}o Paulo, FAPESP, which has supported the present research through a doctoral scholarship to the first author under Grant No.\ 2021/00147-8\@. Additional support from FAPESP under the Research Grant No.\ 2013/07375-0 is also greatly appreciated. The authors also gratefully acknowledge the support for the present research provided by Conselho Nacional de Desenvolvimento Científico e Tecnológico, CNPq, under the Research Grant No.\ 309985/2013-7\@.

\section*{REFERENCES} 
\label{Sec:references}

\bibliographystyle{AIAAbst}
\renewcommand{\refname}{}
\bibliography{bibfile}

\begin{thebibliography}{10}
\newcommand{\enquote}[1]{``#1''}

\bibitem{breviglieri2016}
Breviglieri, C., {\em A Portable Parallel Computing Framework for Aerospace Simulations\/}, Ph.D. thesis, Instituto Tecnológico de Aeronáutica, São José dos Campos, SP, Brazil, 2016.

\bibitem{wang2012}
{Wang, Z. J. \emph{et. al.}}, \enquote{High-Order {CFD} Methods: Current Status and Perspective,} {\em International Journal for Numerical Methods in Fluids\/}, Vol.~43, 2012, pp.~1--42.

\bibitem{wang2007}
Wang, Z.~J., \enquote{High-Order Methods for the {E}uler and {N}avier-{S}tokes Equations on Unstructured Grids,} {\em Progress in Aerospace Sciences\/}, Vol.~43, 2007, pp.~1--41.

\bibitem{ekaterinaris2005}
Ekaterinaris, J.~A., \enquote{High-Order Accurate, Low Numerical Diffusion Methods for Aerodynamics,} {\em Progress in Aerospace Sciences\/}, Vol.~41, 2005, pp.~192--300.

\bibitem{huynh2014}
Huynh, H.~T., Wang, Z.~J., and Vincent, P.~E., \enquote{High-Order Methods for Computational Fluid Dynamics: A Brief Review of Compact Differential Formulation on Unstructured Grids,} {\em Computers \& Fluids\/}, Vol.~98, 2014, pp.~209--220.

\bibitem{cockburn1991}
Cockburn, B. and Shu, C.~W., \enquote{The {R}unge-{K}utta Local Projection P{$^1$}-Discontinuous-{G}alerkin Finite Element Method for Scalar Conservation Laws,} {\em M{$^2$}AN Mathematical Modelling and Numerical Analysis\/}, Vol.~25, No.~3, 1991, pp.~337--361.

\bibitem{bassi1997}
Bassi, F. and Rebay, S., \enquote{A High-Order Accurate Discontinuous Finite Element Method for the Numerical Solution of the Compressible {N}avier-{S}tokes Equations,} {\em Journal of Computational Physics\/}, Vol.~131, No.~2, 1997, pp.~267--279.

\bibitem{liu2006}
Liu, Y., Vinokur, M., and Wang, Z.~J., \enquote{Discontinuous Spectral Difference Method for Conservation Laws on Unstructured Grids,} {\em Journal of Computational Physics\/}, Vol.~216, 2006, pp.~780--801.

\bibitem{sun2006}
Sun, Y., Wang, Z.~J., and Liu, Y., \enquote{Spectral (Finite) Volume Method for Conservation Laws on Unstructured Grids {VI}: Extension to Viscous Flow,} {\em Journal of Computational Physics\/}, Vol.~215, No.~1, 2006, pp.~41--58.

\bibitem{huynh2007}
Huynh, H.~T., \enquote{A Flux Reconstruction Approach to High-Order Schemes Including Discontinuous {G}alerkin Methods,} AIAA Paper No.\ 2007-4079, \emph{18th AIAA Computational Fluid Dynamics Conference}, Miami, FL, USA, June 2007.

\bibitem{huynh2009}
Huynh, H.~T., \enquote{A Reconstruction Approach to High-Order Schemes Including Discontinuous {G}alerkin for Diffusion,} AIAA Paper No.\ 2009-403, \emph{47th AIAA Aerospace Sciences Meeting Including The New Horizons Forum and Aerospace Exposition}, Orlando, FL, USA, Jan. 2009.

\bibitem{roe1986}
Roe, P.~L., \enquote{Characteristic-based schemes for the {E}uler Equations,} {\em Annual Review of Fluid Mechanics\/}, Vol.~18, 1986, pp.~337--365.

\bibitem{kopriva1996}
Kopriva, D.~A. and Kolias, J.~H., \enquote{A Conservative Staggered-Grid {C}hebyshev Multidomain Method for Compressible Flows,} {\em Journal of Computational Physics\/}, Vol.~125, 1996, pp.~244--261.

\bibitem{lomax2001}
Lomax, H., Pulliam, T.~H., and Zingg, D.~W., {\em Fundamentals of Computational Fluid Dynamics\/}, Springer-Verlag, Heidelberg, 1st ed., 2001, p. 249.

\bibitem{romero2016}
Romero, J., Asthana, K., and Jameson, A., \enquote{A Simplified Formulation of the Flux Reconstruction Method,} {\em Journal of Scientific Computing\/}, Vol.~67, 2016, pp.~351--374.

\bibitem{hirschV1}
Hirsch, C., {\em Numerical Computation of Internal and External Flows - Volume 1: Fundamentals of Numerical Discretization\/}, John Wiley \& Sons, Chichester, 1st ed., 1988, p. 515.

\bibitem{roe2017}
Roe, P., \enquote{A Simple Explanation of Superconvergence for Discontinuous {G}alerkin Solutions to $u_t + u_x = 0$,} {\em Communications in Computational Physics\/}, Vol.~21, No.~4, 2017, pp.~905--912.

\end{thebibliography}

\section*{RESPONSIBILITY NOTICE}

The authors are solely responsible for the printed material included in this paper.

\end{document}